\documentclass[
aps,
prb,
reprint, 
twocolumn,
groupedaddress,
superscriptaddress,
amsfonts,
footinbib,
floatfix,
longbibliography
]{revtex4-2}

\RequirePackage{amsmath,amssymb,bm,bbm}
\RequirePackage{graphicx}
\RequirePackage[usenames,dvipsnames]{xcolor}
\usepackage{float}

\usepackage{pdfpages}
\makeatletter\AtBeginDocument{\let\LS@rot\@undefined}\makeatother

\usepackage[T1]{fontenc}

\usepackage{enumitem}
\usepackage{verbatim}

\usepackage{tabularx}

\usepackage{hyperref}
\hypersetup{
    colorlinks=true,
    citecolor= blue,
    linkcolor= blue,
    filecolor=blue,      
    urlcolor= blue,
    pdfcreator = {\LaTeX\ and \flqq hyperref\frqq},
}

\usepackage[capitalise]{cleveref}
\crefname{section}{Sec.}{Secs.}

\graphicspath{ {./Figs/} }

\newcommand{\Appref}[1]{Appendix~\ref{#1}}

\newcommand{\bbZ}{\mathbb{Z}}
\newcommand{\K}{\mathrm{K}}
\newcommand{\M}{\mathrm{M}}
\newcommand{\bk}{\mathbf{k}}
\newcommand{\bq}{\mathbf{q}}
\newcommand{\bp}{\mathbf{p}}
\newcommand{\bQ}{\mathbf{Q}}
\newcommand{\bb}{\mathbf{b}}
\newcommand{\br}{\mathbf{r}}

\newcommand{\ba}{\mathbf{a}}
\newcommand{\bc}{\mathbf{c}}
\newcommand{\bd}{\mathbf{d}}
\newcommand{\bepsilon}{\bm{\epsilon}}
\newcommand{\bh}{\mathbf{h}}
\newcommand{\bdelta}{\bm{\delta}}
\newcommand{\scT}{\mathcal{T}}
\newcommand{\scM}{\mathcal{M}}
\newcommand{\scR}{\mathcal{R}}
\newcommand{\fourv}{\text{4v}}
\newcommand{\twov}{\text{2v}}

\makeatletter
\newcommand*{\rom}[1]{\expandafter\@slowromancap\romannumeral #1@}
\makeatother

\renewcommand{\Re}{\operatorname{Re}}
\renewcommand{\Im}{\operatorname{Im}}

\DeclareMathOperator{\sgn}{sgn}

\begin{document}

\title{Dynamical Response of the Kitaev Spin Liquid under Third-Nearest-Neighbor Heisenberg Interaction}

\author{Chuan Chen}
\affiliation{School of Physical Science and Technology,
Lanzhou University,
Lanzhou 730000, China}
\affiliation{Lanzhou Center for Theoretical Physics, 
Key Laboratory of Quantum Theory and Applications of MoE,
Key Laboratory of Theoretical Physics of Gansu Province,
Gansu Provincial Research Center for Basic Disciplines of Quantum Physics,
Lanzhou University,
Lanzhou 730000, China}

\author{Jiucai Wang}
\email[Correspondence to: ]{jcwangphys@hznu.edu.cn}
\affiliation{School of Physics,
Hangzhou Normal University,
Hangzhou 311121, China}

\begin{abstract}
Motivated by growing evidence for the significance of the third-nearest-neighbor Heisenberg ($J_3$) interaction in candidate Kitaev materials,
we investigate the dynamical properties of the Kitaev spin liquid (KSL) under a $J_3$ perturbation, focusing on its spin dynamical structure factor (DSF) 
and Raman scattering.
Within a self-consistent parton mean-field plus random-phase approximation framework, we find that $J_3$ induces coherent, 
paramagnon-like collective modes that coexist with a high-energy Majorana continuum in the spin DSF.
The softening of these modes with increasing $|J_3|$ signals a quantum phase transition to magnetic order. 
Remarkably, magnetic ordering sets in at a common critical $J_3$ for both ferromagnetic ($K<0$) and antiferromagnetic ($K>0$) Kitaev models, 
with the resulting ordered states forming exact dual pairs under a four-sublattice duality transformation that maps $(K,J_3) \rightarrow (-K,J_3)$.
An external magnetic field further softens the preexisting paramagnon modes, thereby enhancing magnetic order.
Perturbative Raman calculations show that while the Kitaev-like Raman vertex probes only
itinerant matter Majorana fermions,
the response from the $J_3$-like vertex features both matter Majoranas and visons.
Four-vison excitations produce a sharp peak accompanied by a two-fermion continuum,
whereas two-vison excitations yield a continuum closely resembling the single-matter-fermion density of states.
These results provide a unified perspective on the dynamical signatures of $J_3$-perturbed KSL and 
are helpful for interpreting experimental spectra in candidate Kitaev materials with sizable $J_3$ interactions.
\end{abstract}

\date{\today}

\maketitle

\section{Introduction}
Quantum spin liquids (QSLs) are exotic phases of matter characterized by long-range entanglement and fractionalized excitations.
They typically emerge in strongly frustrated quantum spin systems, where geometric frustration or quantum fluctuations
preclude the formation of long-range magnetic order~\cite{Lee2006-mw,Balents2010-td,Savary2017-bs,Zhou2017-pe,Broholm2020-pq}.
A paradigmatic example is the exactly solvable Kitaev honeycomb model~\cite{Kitaev2006-xm}, in which bond-dependent interactions
stabilize a $\bbZ_2$ QSL hosting itinerant matter Majorana fermions coupled to static $\bbZ_2$ gauge-flux excitations (visons).
A key step toward materials realization was taken by Jackeli and Khaliullin~\cite{Jackeli2009-yn},
who showed that such anisotropic interactions can naturally arise in spin-orbit-coupled transition-metal compounds.
This insight sparked extensive efforts to identify and investigate candidate Kitaev materials over the past decade. 
Prominent examples include the iridates A$_2$IrO$_3$ (A=Li, Na), $\alpha$-RuCl$_3$, and cobaltates such as Na$_2$Co$_2$TeO$_6$
(for reviews, see Refs.~\cite{Hermanns2018-vh,Knolle2019-hn,Takagi2019-ok,Trebst2022-go,Rousochatzakis2024-td,Matsuda2025-pd}).
In practice, however, these materials host substantial non-Kitaev interactions---most notably nearest-neighbor (NN) Heisenberg and
off-diagonal $\Gamma$ and $\Gamma'$ couplings---which typically drive magnetic ordering at low temperatures.
An external magnetic field can suppress the magnetic order and drive the system into a quantum-disordered regime prior to
full spin polarization at high fields. 
For example, $\alpha$-RuCl$_3$ exhibits a field-induced disordered phase with signatures suggestive of a proximate QSL,
although its precise nature remains under active debate~\cite{Banerjee2016-nn,Banerjee2017-hq,Banerjee2018-km,Kasahara2018-or,Yokoi2021-ul,Czajka2021-tm,Czajka2023-da,Lefrancois2022-dl,Lefrancois2023-wb,Villadiego2021-hu,Chern2021-gc}.

To determine whether a particular QSL is realized in a perturbed Kitaev material, it is essential to identify the underlying microscopic
spin model and clarify how non-Kitaev interactions affect the Kitaev spin liquid (KSL).
While substantial progress has been made in understanding the effects of NN non-Kitaev interactions, such as Heisenberg and off-diagonal
$\Gamma$ and $\Gamma'$ couplings~\cite{Chaloupka2010-xs,Chaloupka2013-cf,Kimchi2011-ug,Knolle2018-om,Rau2014-ao,Gordon2019-iu,Sorensen2021-jg,Lee2020-zx,Wang2019-bz,Wang2024-yg,Zhang2021-xi,Chuan2025-qb,Maksimov2020-up,Moller2025-ly}, 
increasing evidence points to an essential role of the third-NN Heisenberg interaction. In several candidate materials,
this term is comparable to, or even exceeds, the NN Heisenberg coupling~\cite{Maksimov2020-up,Choi2012-rp,Singh2012-fm,Kim_2022,Kim2020-px,Lin2021-tm,Yao2022-cj}.
For instance, it is crucial for stabilizing the zigzag order in ferromagnetic (FM) Kitaev systems such as $\alpha$-RuCl$_3$~\cite{Maksimov2020-up}, 
and for reproducing experimentally observed spin dynamical signatures in iridates and cobaltates~\cite{Choi2012-rp,Kim2020-px,Lin2021-tm,Kim_2022,Songvilay_2020,Samarakoon_2021,Sanders_2022,osti_2406719,Yao2022-cj}.
These observations motivate a systematic investigation of the impact of third-NN Heisenberg interaction on the KSL.

In this work, we study the Kitaev-$J_3$ ($K$-$J_3$) model, which comprises Kitaev ($K$) and third-NN Heisenberg ($J_3$) interactions.
We focus in particular on how the $J_3$ term modifies the signatures of the KSL 
in two experimentally relevant dynamical probes:
the spin dynamical structure factor (DSF) and Raman scattering.

First, we compute the spin DSF using a recently developed self-consistent parton mean-field plus 
random-phase approximation (RPA) framework~\cite{Rao2025-bg}.
We find that the $J_3$ interaction induces low-energy paramagnon-like collective modes below the Majorana continuum.
The excitation gaps of these modes collapse at a critical $J_3$, signaling the onset of long-range magnetic order, whose ordering pattern
can be inferred from the momenta of the soft modes.
Specifically, in the FM Kitaev model, the KSL evolves into a zigzag phase for large positive $J_3$, whereas for large negative $J_3$, 
the soft-mode pattern suggests competing tendencies toward ferromagnetic and stripe order (hereafter referred to as FM+stripe regime).
In the AFM Kitaev model, stripe order emerges for large negative $J_3$, while for large positive $J_3$,
the soft modes suggest competing tendencies toward N\'eel AFM and zigzag order (referred to as AFM+zigzag regime).
In both the FM+stripe and AFM+zigzag regimes, the soft modes may correspond either to nearly degenerate single-$\bQ$ orders---where a single wavevector is ultimately selected---or to multi-$\bQ$ states, depending on the underlying energetics.
Importantly, the AFM+zigzag (FM+stripe) ordering tendencies are related to the zigzag (stripe) phases through
a duality transformation of the model that maps $K \rightarrow -K$.
We further investigate the effect of an external magnetic field applied along different crystallographic directions and
find that it consistently enhances the tendency toward magnetic ordering.

Second, we study Raman scattering using perturbation theory combined with the exact solution of the Kitaev model.
Although the pure Kitaev interaction yields a two-Majorana-fermion continuum~\cite{Knolle2014-ma,Nasu2016-sk}, 
the $J_3$ term introduced two additional contributions.
One arises from intermediate four-vison excitations and resembles the response of the NN Heisenberg interaction~\cite{Knolle2014-ma},
while the other originates from intermediate two-vison excited states. 
Interestingly, the Raman response associated with the latter process mimics the single-particle density of states of the itinerant matter
Majorana fermions, even though a local probe cannot excite an isolated matter fermion.

The remainder of this paper is organized as follows.
In \cref{sec:model}, we introduce the $K$-$J_3$ model and discuss its symmetries.
In \cref{sec:DSF}, we present the mean-field analysis of the KSL
and compute the spin DSF within a mean-field plus RPA framework,
covering different $(K,J_3)$ regimes and the effect of an external magnetic field. 
The Raman response of the $J_3$-perturbed KSL is analyzed in \cref{sec:Raman}.
Finally, we conclude with a discussion in \cref{sec:discuss}.
Additional technical details are provided in the Appendices.

\begin{figure}[htbp]
\centering
\includegraphics[width=0.5 \textwidth]{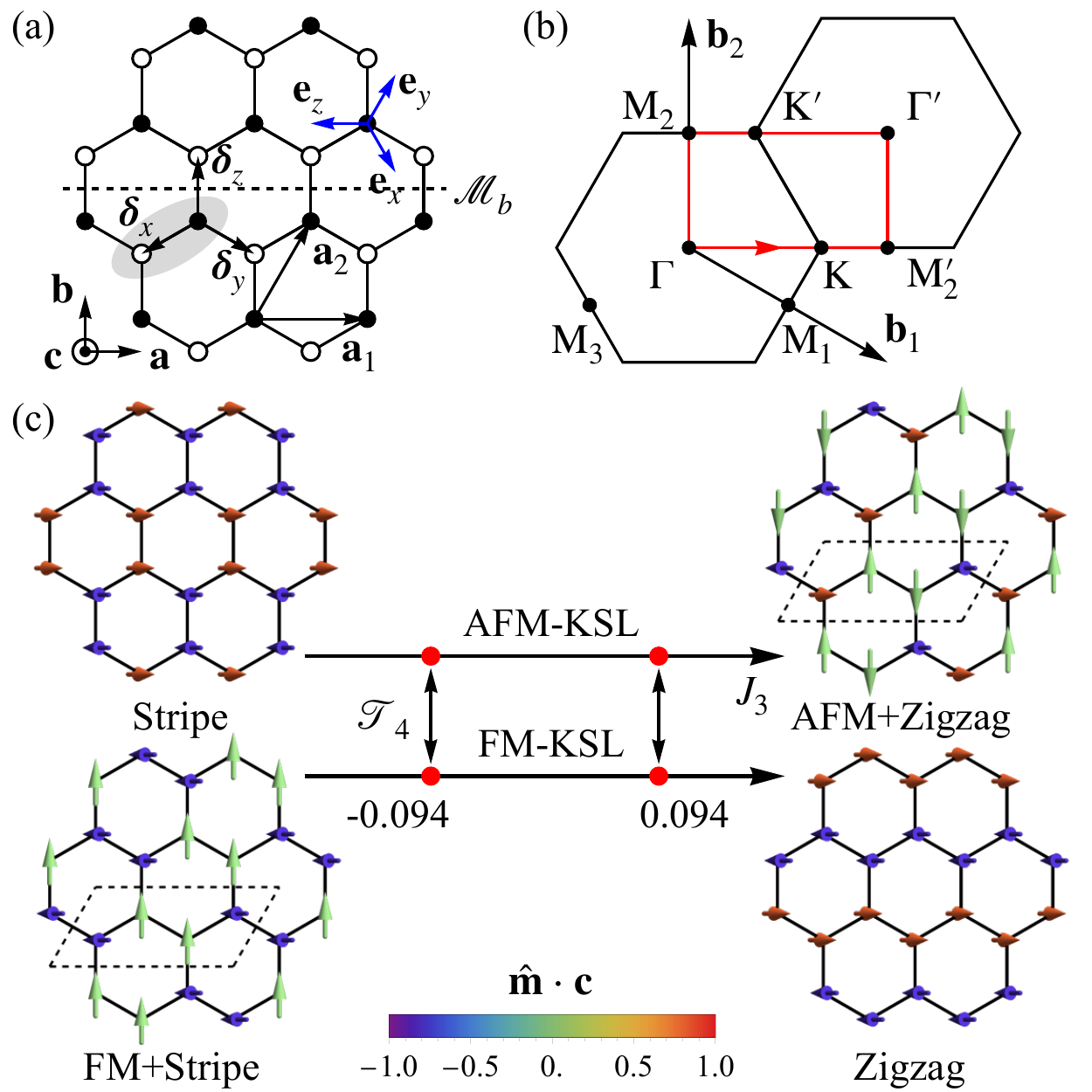}
\caption{
(a) Schematic of the honeycomb lattice and spin axes.
The A and B sublattices are indicated by black and white dots, respectively. The gray shaded region indicates a unit cell.
The $\alpha$-type bonds are aligned parallel to $\bdelta_\alpha$.
The spin axes are illustrated by blue arrows labeled $\mathbf{e}_\alpha$.
The crystal axes $\mathbf{a}$, $\mathbf{b}$, and $\mathbf{c}$ are also shown.
The dashed line indicates the mirror plane associated with the transformation $\scM_b$.
(b) Brillouin zone and high-symmetry $k$ points.
The red contour indicates the momentum path used to plot the spin DSF.
(c) Magnetic orders induced by the $J_3$ interaction in both the FM ($K=-1$) and AFM ($K=1$) Kitaev models.
For the FM Kitaev interaction, zigzag order emerges when $J_3 \ge 0.094$,
while for $J_3 \le -0.094$, the soft modes indicate competing tendencies toward FM and stripe order (FM+stripe).
For the AFM Kitaev interaction, stripe order emerges when $J_3 \le -0.094$, whereas for $J_3 \ge 0.094$,
the soft modes suggest competing tendencies toward N\'eel AFM and zigzag order (AFM+zigzag).
The critical values of $J_3$ coincide in the two models due to the $\scT_4$ duality transformation $K \rightarrow -K$,
under which the corresponding states are related.
Note that the zigzag (stripe) configuration illustrated here is shown with equal spin-$x$ and spin-$y$ components as an example; 
under the $\scT_4$ duality, this maps to a state with spin $x$ AFM (FM) and spin $y$ zigzag (stripe) components.
This example is intended solely to illustrate the mapping and does not imply that the corresponding phase is necessarily a multi-$\bQ$ state.
The spin $z$ component is set to zero based on energetic considerations.
}\label{fig:schematic}
\end{figure}

\section{The model and its symmetries}\label{sec:model}
The $K$-$J_3$ model is defined on a honeycomb lattice, with NN Kitaev ($K$) and third-NN Heisenberg ($J_3$) interactions:
\begin{align}\label{eq:H_K-J3}
    H = \sum_{\langle i,j \rangle_\alpha} K \, \sigma_i^\alpha \sigma_j^\alpha
    + \sum_{\langle \! \langle \! \langle i,j \rangle \! \rangle \! \rangle} J_3 \, \bm{\sigma}_i \cdot \bm{\sigma}_j - \sum_{i} \mathbf{h} \cdot \bm{\sigma}_i.
\end{align}
Here $\langle i,j \rangle_\alpha$ denotes a NN $\alpha$-bond ($\alpha = x,y,z$),
while $\langle \! \langle \! \langle i,j \rangle \! \rangle \! \rangle$ denotes a third-NN bond.
A Zeeman term is also included to account for the effect of an external magnetic field.
In this work, we consider magnetic fields applied along the three crystallographic axes $\mathbf{a}$, $\mathbf{b}$, and $\mathbf{c}$.
A schematic of the honeycomb lattice and spin axes is shown in \cref{fig:schematic}(a).

The model exhibits several symmetries that simplify the mean-field analysis by reducing the number of independent mean-field parameters.
In the absence of a magnetic field, the model is invariant under the following symmetry operations (c.f. \cref{fig:schematic}(a)):
(i)~lattice translations $T_1$ and $T_2$ along the primitive vectors $\ba_1$ and $\ba_2$, respectively;
(ii)~a sixfold rotation about the $c$ axis followed by a mirror reflection with respect to the $a$-$b$ plane (denoted by $C_6$);
(iii)~time-reversal symmetry $\scT$; and
(iv)~mirror reflection across the plane perpendicular to the $b$-axis (denoted by $\scM_b$).

When an external magnetic field is applied, the symmetry of the model is reduced.
The $C_6$ rotation is preserved when $\bh \parallel \bc$, whereas the mirror symmetry $\scM_b$
remains intact for $\bh \parallel \bb$.
Notably, for $\bh \parallel \ba$ and $\bh \parallel \bc$, although both $\scT$ and $\scM_b$ are individually broken,
their product $\scT \scM_b$ remains a symmetry of the system~\cite{Zou2020-ni}.
A complete list of the model's symmetries for fields applied along different crystallographic directions is given in~\cref{tab:symmetry}.
\begin{table}[t]
    \centering
    \caption{Symmetries of the $K$-$J_3$ model for magnetic fields applied along different crystallographic directions.}
    \label{tab:symmetry}
    \begin{tabular*}{\linewidth}{@{\extracolsep{\fill}} c c c c c c}
      \hline \hline
             & $T_{1,2}$ & $C_6$ & $\scM_b$ & $\scT$ & $\scT \scM_b$  \\
      \hline
      $\bh = 0$  & \checkmark & \checkmark & \checkmark & \checkmark & \checkmark \\
      $\bh \parallel \ba$  & \checkmark & $\times$ & $\times$ & $\times$ & \checkmark \\
      $\bh \parallel \bb$  & \checkmark & $\times$ & \checkmark & $\times$ & $\times$ \\
      $\bh \parallel \bc$  & \checkmark & \checkmark & $\times$ & $\times$ & \checkmark \\
      \hline \hline
    \end{tabular*}
\end{table}

Besides these symmetries, the model also possesses an exact duality ($\scT_4$), a four-sublattice unitary transformation
that maps $(K,J_3) \rightarrow (-K,J_3)$ (see \Appref{sec:T_4} for its explicit form)~\cite{Rousochatzakis2024-td}.
This implies that the FM and AFM KSLs undergo phase transitions at the same critical value of $J_3$,
with the resulting phases related by the $\scT_4$ transformation.
This further indicates that the FM and AFM KSLs share a common robustness against the $J_3$ interaction,
unlike their differing repsonse to other non-Kitaev interactions. 
For example, the FM (AFM) KSL is more stable against the NN Heisenberg ($\Gamma$) interaction.
As will be shown below, the phase diagram obtained from our spin DSF calculations
is fully consistent with this property of the model.

\begin{figure*}[htbp]
\centering
\includegraphics[width = \textwidth]{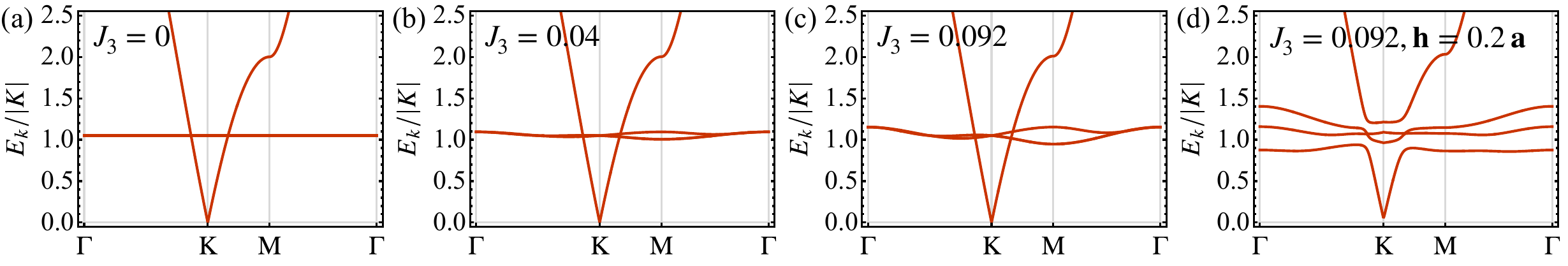}
\caption{Mean-field fermion band dispersions at selected $J_3$ for an FM Kitaev interaction ($K=-1$).
(a) Fermion bands for the pure FM Kitaev model ($J_3 = 0$). The gauge fermions are static and form degenerate flat bands.
(b) A finite $J_3$ term ($J_3 = 0.04$) induces gauge fluctuations, rendering the gauge-fermion bands dispersive.
(c) The gauge-fermion bands become more dispersive at larger $J_3 = 0.092$.
(d) Fermion bands at $J_3 = 0.092$ in the presence of a magnetic field $\bh=0.2\, \ba$.
Hybridization between gauge and matter fermions further enhances the band dispersion and gaps out the Dirac cone at the $\K$ point.
At zero field, reflection-type symmetries along certain high-symmetry lines interchange pairs of crystal-symmetry-related gauge-fermion flavors,
leading to pairwise degeneracies of the corresponding bands. In a finite magnetic field, the matter and gauge fermions become coupled, 
and these degeneracies are generally lifted.
}\label{fig:MF-bands}
\end{figure*}

\section{Spin dynamical structure factor from mean-field plus RPA}\label{sec:DSF}
The spin DSF, measurable via inelastic neutron scattering and resonant inelastic X-ray scattering experiments,
encodes the dynamical properties of a magnetic system and provides a direct window into the fractionalized excitations (e.g., spinons)
of QSLs~\cite{Savary2017-bs,Knolle2019-hn}.
While the DSF can be computed exactly for the pure Kitaev model~\cite{Knolle2014-wl,Knolle2015-lx}, 
non-Kitaev interactions in real materials render an exact evaluation intractable.
We therefore employ a recently developed self-consistent parton mean-field plus RPA framework~\cite{Rao2025-bg},
previously shown to qualitatively capture the ordered phases induced by NN non-Kitaev couplings such as Heisenberg and $\Gamma$ interactions,
and apply it here to compute the spin DSF of the $J_3$-perturbed KSL.

\begin{figure*}
\centering
\includegraphics[width=0.85 \textwidth]{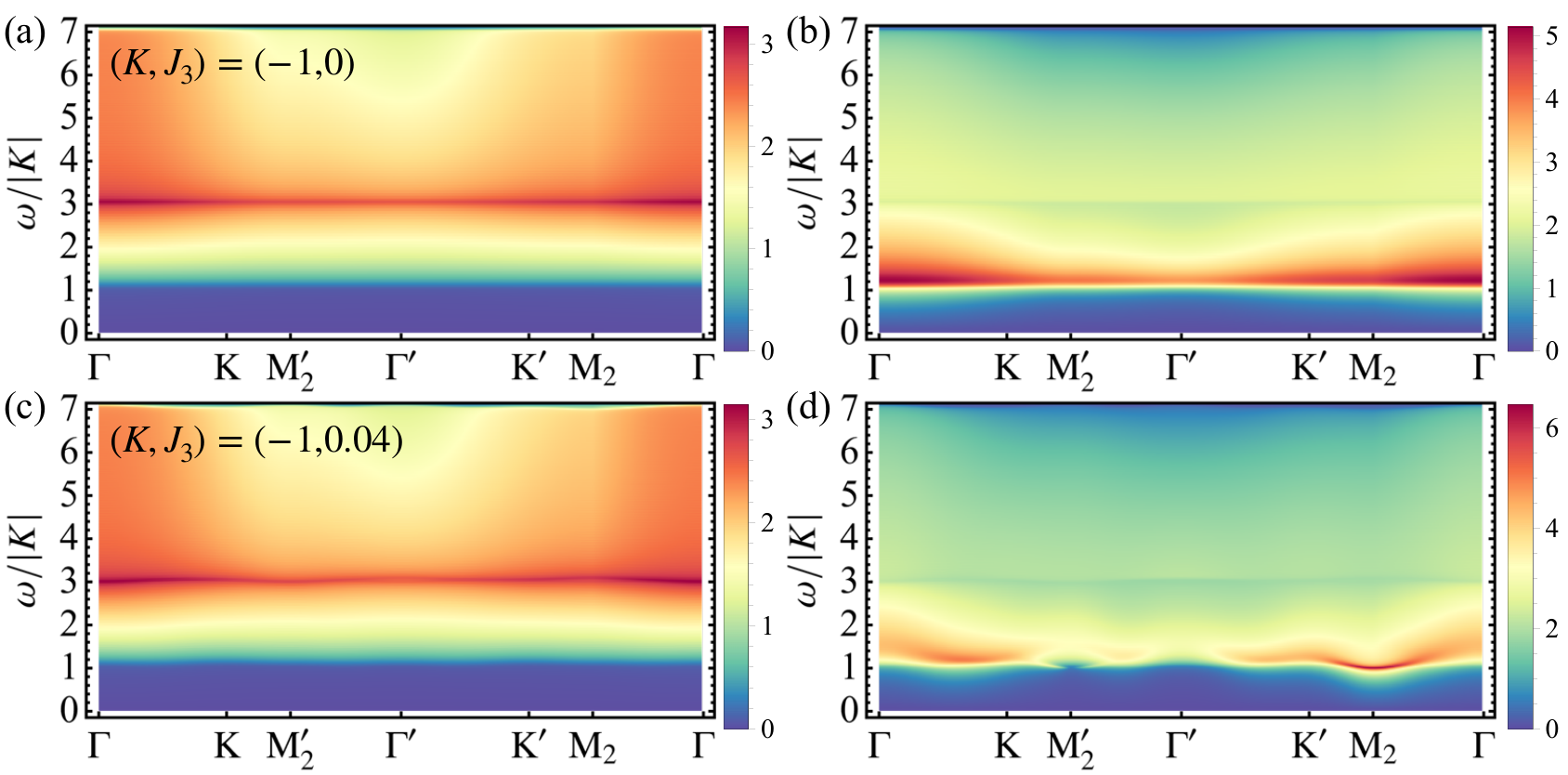}
\caption{Mean-field [(a),~(c)] versus RPA-corrected [(b),~(d)] spin DSFs: 
(a)--(b) for the ferromagnetic Kitaev model $(K,J_3)=(-1,0)$; (c)--(d) for $(K,J_3)=(-1,0.04)$.
The structure factors are plotted on a logarithmic scale, $\ln[1+S(\omega,\bq)]$.
}\label{fig:MF-vs-RPA}
\end{figure*}

\subsection{Mean-field theory of the KSL}\label{sec:mean-field}
The mean-field theory of the KSL is most conveniently formulated using the Majorana representation of the spin operators,
in which a local spin operator is expressed as a bilinear of Majorana fermions,
$\sigma_j^\mu \leftrightarrow i \gamma_{j,\mu} \gamma_{j,0}$, subject to the constraint
$\gamma_{j,x} \gamma_{j,y} \gamma_{j,z} \gamma_{j,0} = 1$.
Here $\gamma_{x,y,z}$ and $\gamma_0$ are equivalent to the $b^{x,y,z}$ and $c$
operators introduced in the original work by Kitaev~\cite{Kitaev2006-xm}.
Under this representation, a two-spin term in the Hamiltonian is mapped onto a four-Majorana interaction, which is subsequently
decoupled into bilinears of Majorana operators within the mean-field approximation.
For example, the Kitaev interaction on a bond $\langle i,j \rangle_\alpha$ (with $i \in A$) is written as
$\sigma_i^\alpha \sigma_j^\alpha \leftrightarrow i \gamma_{i,\alpha} \gamma_{i,0} \, i \gamma_{j,\alpha} \gamma_{j,0}$,
which is decoupled as
\begin{align}\label{eq:decouple-Kitaev}
    & u_{\alpha,\alpha} \, i \gamma_{i,0} \gamma_{j,0} + u_{\alpha,0} \, ( -i \gamma_{i,\alpha} \gamma_{j,\alpha} ) - u_{\alpha,\alpha} u_{\alpha,0} \nonumber \\
    & + m_\alpha( i\gamma_{i,\alpha} \gamma_{i,0} + i \gamma_{j,\alpha} \gamma_{j,0} ) - m_\alpha^2.
\end{align}
The mean-field parameters above are determined self-consistently as
\begin{subequations}
\begin{align}
    & u_{\alpha,\alpha} \equiv \langle -i \gamma_{i,\alpha} \gamma_{j,\alpha} \rangle, \,
    u_{\alpha,0} \equiv \langle i \gamma_{i,0} \gamma_{j,0} \rangle, \label{eq:u_alpha}  \\
    & m_\alpha \equiv \langle i \gamma_{i,\alpha} \gamma_{i,0} \rangle = \langle i \gamma_{j,\alpha} \gamma_{j,0} \rangle \label{eq:m}.
\end{align}    
\end{subequations}
Here we have invoked the translational symmetry of the KSL and assumed a uniform magnetization throughout the system.

Similarly, for the Heisenberg interaction on a third-NN bond $\langle\!\langle\!\langle i,j \rangle\!\rangle\!\rangle \in \alpha$
(with $i \in A$ and the bond oriented along the $\alpha$ direction), 
$\bm{\sigma}_i \cdot \bm{\sigma}_j$ is decoupled as
\begin{align}\label{eq:decouple-J3}
    \sum_{\mu = x,y,z} & \left[ v_{\alpha,\mu} \, i \gamma_{i,0} \gamma_{j,0} + v_{\alpha,0} (-i \gamma_{i,\mu} \gamma_{j,\mu}) - v_{\alpha,\mu} v_{\alpha,0} 
    \right. \nonumber \\
    & \left. + m_\mu (i\gamma_{i,\mu} \gamma_{j,0} + i \gamma_{j,\mu} \gamma_{j,0} ) - m_{\mu}^2 \right].
\end{align}
The parameters $v_{\alpha,\mu}$ and $v_{\alpha,0}$ are determined through:
\begin{equation}
    v_{\alpha,\mu} \equiv \langle -i \gamma_{i,\mu} \gamma_{j,\mu} \rangle, \,
    v_{\alpha,0} \equiv \langle i \gamma_{i,0} \gamma_{j,0} \rangle \label{eq:v_alpha}.
\end{equation}

Finally, the constraint $\gamma_{x} \gamma_{y} \gamma_{z} \gamma_{0} = 1$, which is equivalent to
$i \gamma_{x} \gamma_0 + i \gamma_{y} \gamma_z = 0$ (and all cyclic permutations of $x$, $y$, and $z$),
is imposed on-average by adding the following term to the mean-field Hamiltonian
\begin{align}\label{eq:Lagrange}
    \sum_i & \, \lambda_x ( i \gamma_{i,x} \gamma_0 + i \gamma_{i,y} \gamma_{i,z} )
    + \lambda_{y} ( i \gamma_{i,y} \gamma_{i,0} + i \gamma_{i,z} \gamma_{i,x} ) \nonumber \\
    & + \lambda_z ( i \gamma_{i,z} \gamma_{i,0} + i \gamma_{i,x} \gamma_{i,y} ).
\end{align}
Here the $\lambda_x$ is a Lagrange multiplier which is determined by $\langle i \gamma_{i,x} \gamma_{i,0} + i \gamma_{i,y} \gamma_{i,z} \rangle = 0$,
while $\lambda_y$ and $\lambda_z$ are fixed analogously.

For the ideal Kitaev model, the KSL ground state satisfies (for $\alpha = x, y, z$):
\begin{equation}
u_{\alpha,\alpha} = 1,\, u_{\alpha,0} \approx -\sgn(K) \times 0.525,\, m_\alpha = \lambda_\alpha = 0.
\end{equation}
The projective symmetry group (PSG)~\cite{Wen2002-zt} of the associated parton mean-field Hamiltonian can then be deduced, which encodes
how the Majorana fermions transform under the various symmetry operations listed in \cref{sec:model} (see Table~I of Ref.~\cite{You2012-gj}).
In this study, we focus on the KSL perturbed by a $J_3$ interaction and an external magnetic field, whose PSG is a subgroup of that of the ideal KSL.
The projective implementations of the remaining symmetries therefore coincide with those in the ideal case.

The symmetry of the mean-field Hamiltonian will impose constraints that reduce the number of independent mean-field parameters.
In the absence of an external magnetic field, the time-reversal symmetry enforces $m_\alpha = \lambda_\alpha = 0$ ($\alpha = x,y, z$),
while the $C_6$ and $\scM_b$ symmetries require
\begin{subequations}
\begin{align}
    & u_{x,0} = u_{y,0} = u_{z,0},\, u_{x,x} = u_{y,y} = u_{z,z}, \\ 
    & v_{x,0} = v_{y,0} = v_{z,0},\, v_{x,x} = v_{y,y} = v_{z,z}, \\
    & v_{x,y} = v_{y,z} = v_{z,x} = v_{x,z} = v_{y,x} = v_{z,y}.
\end{align}
\end{subequations}
As a result, only five independent mean-field parameters remain to be determined self-consistently.

The mean-field Hamiltonian can be diagonalized in momentum space by defining
\begin{equation}
    \gamma_{\bk,A/B,\mu} = \frac{1}{\sqrt{2 N}} \sum_{\br \in A} e^{-i \bk \cdot \br} \gamma_{\br,A/B,\mu},
\end{equation}
where $N$ is the number of unit cells in the system.
We adopt the convention of labeling a B site (located at $\br+\bdelta_x$) by the coordinate $\br$ of the A site in the same unit cell
(see \cref{fig:schematic}(a)).
The mean-field Hamiltonian is then written as
\begin{align}
    H = \frac{1}{2} \sum_{\bk} 
    \begin{pmatrix}
        \Gamma_{\bk,A}^\dagger, \Gamma_{\bk,B}^\dagger 
    \end{pmatrix}
    h(\bk) 
    \begin{pmatrix}
        \Gamma_{\bk,A} \\
        \Gamma_{\bk,B}
    \end{pmatrix} + \text{const.},
\end{align}
where $\Gamma_{\bk,a}^\dagger = (\gamma_{\bk,a,0}^\dagger, \gamma_{\bk,a,x}^\dagger, \gamma_{\bk,a,y}^\dagger, \gamma_{\bk,a,z}^\dagger )$,
($a = A, B$).
Since $h(-\bk) = -h(\bk)^T = -h(\bk)^*$, the Hamiltonian can be diagonalized as
\begin{align}
    H = \sum_{\bk} \sum_{n = 1}^4 E_{n,\bk} \, \alpha_{\bk,n}^\dagger \alpha_{\bk,n} + \text{const},
\end{align}
where $E_{n,\bk} \ge 0$. The quasiparticles are defined as
\begin{subequations}
\begin{align}
    (\alpha_{\bk,1}^\dagger, \dots ,\alpha_{\bk,4}^\dagger) & = ( \Gamma_{\bk,A}^\dagger, \Gamma_{\bk,B}^\dagger ) u(\bk), \\
    h(\bk) u(\bk) & = u(\bk)
    \begin{pmatrix}
        E_{1,\bk} &  & \\
         & \ddots &  \\
         &  & E_{4,\bk}
    \end{pmatrix}.
\end{align}
\end{subequations}

The fermion band dispersions for an FM Kitaev interaction at selected $J_3$ values are shown in \cref{fig:MF-bands}.
For the pure Kitaev model, the gauge fermions have flat bands, reflecting the fact that
the vison excitations are gapped and static.
As $J_3$ is turned on, these bands become dispersive, indicating that the $J_3$ term induces gauge fluctuations.
In the absence of an external magnetic field, the itinerant Majorana fermions $\gamma_{0}$ are decoupled from 
the gauge fermions $\gamma_{x,y,z}$, and the Dirac cones of the itinerant Majorana fermions are preserved.
In contrast, when $\bh \neq 0$, the hybridization between the two types of fermions
can gap the spectrum.
An analysis of the mean-field Hamiltonian in the presence of a magnetic field is presented in \Appref{sec:MF-with-B}.

\begin{figure*}
\centering
\includegraphics[width=0.85 \textwidth]{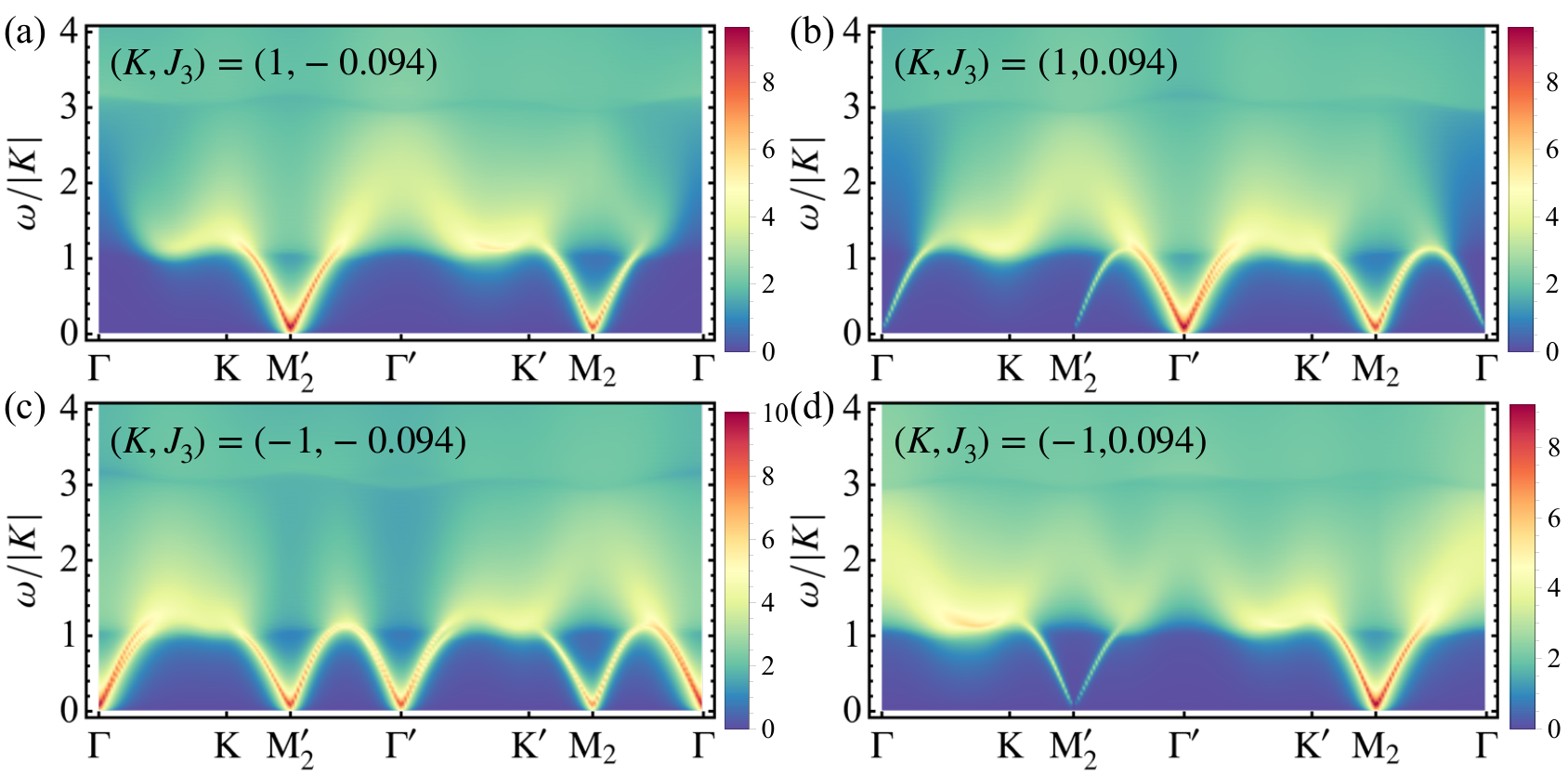}
\caption{RPA spin DSFs in different regimes of $(K,J_3)$.
For the AFM Kitaev model, the gap closing of the $\M'$ and $\M$ modes at (a) $J_3 = -0.094$ signals a transition to the stripe order,
while the gap closing of the $\Gamma'$ and $\M$ modes at (b) $J_3 = 0.094$ indicates the emergence of the AFM+zigzag regime.
For the FM Kitaev model, at (c) $J_3 = -0.094$, the gap closing of a sharp $\Gamma$ mode together with additional closings at $\Gamma'$, $\M'$, and $\M$
signals a transition to the FM+stripe regime, whereas the ``condensation'' of the $\M$ mode at (d) $J_3 = 0.094$ indicates a transition to the zigzag order.
}\label{fig:S-orders}
\end{figure*}

\subsection{RPA calculation of the spin susceptibility}
The spin DSF is defined as
\begin{equation}\label{eq:S-fac}
    S_{\mu,\nu}(\omega,\bq) = \frac{1}{N} \int dt \, e^{i \omega t} \langle \sigma^\mu_{\bq}(t) \sigma^\nu_{-\bq}(0) \rangle.
\end{equation}
According to the fluctuation-dissipation theorem~\cite{Auerbach1998-rx}, it is related to the spin susceptibility through
\begin{align}
    S_{\mu,\nu}(\omega,\bq) & = \frac{-2}{1-e^{-\beta \omega}}\Im \chi_{\mu,\nu}(i\Omega_n \rightarrow \omega+i \eta,\bq),
\end{align}
where the prefactor reduces to $-2$ in the zero-temperature limit and for $\omega > 0$.
The imaginary-time spin susceptibility can be written as
\begin{align}
    \chi_{\mu,\nu} = (1, e^{-i \bq \cdot \bdelta_x})
    \begin{pmatrix}
        \chi_{A \mu,A \nu} & \chi_{A \mu,B \nu} \\
        \chi_{B \mu,A \nu} & \chi_{B \mu,B \nu}
    \end{pmatrix}
     \begin{pmatrix}
         1 \\
         e^{i \bq \cdot \bdelta_x}
     \end{pmatrix},
\end{align}
where the argument $(i\Omega_n,\bq)$ has been suppressed for brevity.
The sublattice-resolved spin susceptibility is defined as
\begin{subequations}
\begin{align}
    \chi_{a \mu, b \nu}(i\Omega_n,\bq) & = \int d\tau \, e^{i\Omega_n \tau} \chi_{a \mu, b \nu}(\tau,\bq), \\
    \chi_{a \mu, b \nu}(\tau,\bq) & = -\frac{1}{N} \langle T \sigma_{\bq,a}^\mu(\tau) \sigma_{-\bq,b}^\nu(0) \rangle.
\end{align}
\end{subequations}
Utilizing the Majorana representation of spins, the spin susceptibility at the mean-field level
reduces to a convolution of two single-particle Green's functions
\begin{align}\label{eq:chi^0}
    & \chi_{a \mu,b \nu}^{0}(i\Omega_n,\bq) \nonumber \\
    & = -\frac{(2i)^2}{N \beta} \sum_{\omega_n, \bk} \left[ G_{a \mu, b \nu}^0(i\omega_n + i\Omega_n,\bk+\bq) G_{b 0, a 0}^0(i\omega_n, \bk) \right. \nonumber \\
    &\quad \left. - G_{a \mu, b 0}^0(i\omega_n + i\Omega_n,\bk+\bq) G_{b \nu, a 0}^0(i\omega_n, \bk) \right],
\end{align}
where the single-particle Green's function is defined as ($\mu, \nu = 0,x,y,z$)
\begin{equation}
    G_{a \mu,b \nu}^0(\tau,\bk) = -\langle T \gamma_{\bk,a,\mu}(\tau) \gamma_{\bk,b,\nu}^\dagger(0) \rangle_0.
\end{equation}
Transforming to Matsubara frequency leads to
\begin{align}\label{eq:G^0}
    & G_{a \mu, b \nu}^0(i \omega_n, \bk) \nonumber \\
    & = \sum_{l=1}^4 \left[ \frac{u(\bk)_{a \mu,l} u(\bk)^\dagger_{l,b \nu}}{i \omega_n - E_{l,\bk}}
    + \frac{ u(-\bk)^*_{a \mu,l} u(-\bk)^T_{l,b \nu} }{i \omega_n + E_{l,\bk}} \right].
\end{align}
\cref{eq:chi^0,eq:G^0} indicate that the mean-field spin DSF forms a two-particle continuum, as shown in \cref{fig:MF-vs-RPA}(a).

To go beyond the mean-field approximation, we adopt the RPA formalism of Ref.~\cite{Rao2025-bg}, which partially incorporates
the fluctuations around the saddle point and provides a quantitatively good approximation to the spin DSF of the ideal Kitaev model.
Within the RPA calculation, the original spin–spin interactions---expressed in terms of Majorana fermions---are reintroduced.
For the $K$-$J_3$ model, the interaction term reads
\begin{align}
    \frac{-(2i)^2}{N} \sum_{\bk,\bk',\bp} \sum_{a,b,\mu,\nu}
    U(\bp)_{a \mu, b \nu} \gamma_{\bk,a,\mu}^\dagger \gamma_{\bk-\bp,a,0} \gamma_{\bk',b,\nu}^\dagger \gamma_{\bk'+\bp,b,0},
\end{align}
where the interaction matrix $U(\bp)$ takes the form
\begin{subequations}
\begin{align}
    & U(\bp) = 
    \begin{pmatrix}
        U(\bp)_{AA} & U(\bp)_{AB} \\
        U(\bp)_{BA} & U(\bp)_{BB}
    \end{pmatrix}, \\
    & U(\bp)_{AB} = U(\bp)_{BA}^\dagger = \frac{-K}{2}
    \begin{pmatrix}
        1 & 0 & 0 \\
        0 & e^{i \bp \cdot \ba_1} & 0 \\
        0 & 0 & e^{i \bp \cdot \ba_2}
    \end{pmatrix} \nonumber \\
    & -\frac{J_3}{2} ( e^{i \bp \cdot (\ba_1+\ba_2)} + e^{i \bp \cdot (\ba_2-\ba_1)} + e^{i \bp \cdot (\ba_1-\ba_2)} ) \, \mathbbm{1}_{3}, \\
    & U(\bp)_{AA} = U(\bp)_{BB} = 0.
\end{align}
\end{subequations}
The RPA susceptibility is related to the mean-field susceptibility in \cref{eq:chi^0} through
\begin{subequations}
\begin{align}
    \chi(i\Omega_n,\bq) & = \left[ 1 + \chi^0(i\Omega_n,\bq) \tilde{U}(\bq) \right]^{-1} \chi^0(i\Omega_n,\bq), \\
    \tilde{U}(\bq) & = U(\bq)+U(-\bq)^T.
\end{align}
\end{subequations}
The spin DSF is then obtained from \cref{eq:S-fac}.
Throughout this paper, we focus on its diagonal components, defined as $S(\omega,\bq) = \sum_{\alpha} S_{\alpha,\alpha}(\omega,\bq)$,
and present them on a logarithmic scale, $\ln[1+S(\omega,\bq)]$.

The RPA correction to the spin dynamics is dramatic.
\cref{fig:MF-vs-RPA} shows the comparison between the mean-field and RPA spin DSFs
for the pure FM Kitaev model $(K,J_3) = (-1,0)$ and for $(K,J_3) = (-1,0.04)$.
In \cref{fig:MF-vs-RPA}(a), the peak in the mean-field structure factor originates from the enhanced
density of states of the itinerant Majorana fermions. 
The RPA correction, however, shifts most of the spectral weight to the bottom of the continuum,
mimicking the effect of local vison excitations on the itinerant Majoranas (see \cref{fig:MF-vs-RPA}(b)).
Remarkably, the RPA susceptibility quantitatively reproduces the exact result of the Kitaev model once the
spectrum is shifted downward to compensate for the mean-field overestimation---by about a factor of four---of the vison-pair excitation energy~\cite{Rao2025-bg}.
In \cref{fig:MF-vs-RPA}(c), the peak in the mean-field structure factor at $J_3 = 0.04$ becomes dispersive, 
reflecting fluctuations of the gauge fermions induced by the $J_3$ interaction (c.f. \cref{fig:MF-bands}(b)).
While the mean-field spectrum remains qualitatively similar to that of the ideal Kitaev model, the RPA-corrected result
differs markedly, as presented in \cref{fig:MF-vs-RPA}(d).
Although the RPA correction still shifts most of the spectral weight to the bottom of the continuum, it now forms a dispersive feature.
A large fraction of the weight accumulates near the band minimum at the $\M$ point, reminiscent of a coherent paramagnon-like collective mode.
The emergence of such collective excitations, together with a featureless higher-energy continuum resembling that of the ideal KSL,
turns out to be a universal feature of the $J_3$-perturbed KSL, independent of the signs of $J_3$ and $K$.
As $J_3$ increases, the dispersive collective excitations gradually separate from the Majorana continuum 
by moving to lower energies and becoming increasingly sharp.
At a critical value of $J_3$, their excitation gaps close at certain high-symmetry momenta, signaling an instability
of the KSL toward long-range magnetic order.
The resulting ordering pattern can be inferred from the momenta of the condensed soft modes
and the associated spectral-weight distribution.

\cref{fig:S-orders} shows the spin DSFs for both FM ($K=-1$) and AFM ($K=1$)
Kitaev interactions at $J_3 = \pm 0.094$.
In the FM Kitaev model, at $J_3 = 0.094$, the sharp collective mode at momentum $\M$ undergoes gap closing, 
indicating a transition to zigzag order.
While the same transition has been reported in previous numerical studies~\cite{Kim2020-px}, the critical value of $J_3$ obtained here differs slightly, 
most likely due to the overestimation of the vison-pair excitation energy within the mean-field treatment~\cite{Knolle2018-om}.
By contrast, in the AFM Kitaev model at the same $J_3$, two sharp modes at the $\Gamma'$ and $\M$ points simultaneously soften and close their gaps,
with the dominant spectral weight at $\Gamma'$.
This reflects competing tendencies toward N\'eel AFM and zigzag order.
Based on the underlying energetics, the system may select either a single-$\bQ$ order (AFM or zigzag) or realize a multi-$\bQ$ state involving both.
We therefore refer to this regime as AFM+zigzag for brevity.
The emergence of the AFM+zigzag regime follows naturally from the four-sublattice duality transformation $\scT_4$, 
which maps $(K,J_3) \rightarrow (-K,J_3)$~\cite{Rousochatzakis2024-td}.
As illustrated in \cref{fig:schematic}(c), applying $\scT_4$ to a zigzag order with wave vector $\M_2$, 
whose ordering moment lies in the $x$-$y$ plane (with the spin $z$ component suppressed due to its energetic unfavorability
under the FM Kitaev interaction), yields a state in which the spin $x$ components form a N\'eel AFM
pattern while the spin $y$ components exhibit a zigzag modulation.
Depending on whether the original zigzag order has one or two finite spin components, 
the $\scT_4$-mapped phase corresponds to a single-$\bQ$ order (AFM or zigzag) or a multi-$\bQ$ state with coexisting AFM and zigzag orders (see \Appref{sec:T_4}).

For $J_3 < 0$, a similar pair of $\scT_4$-related phases emerges.
In the AFM Kitaev model at $J_3 = -0.094$, two sharp modes at $\M'$ and $\M$ close their gaps, with stronger spectral weight at $\M'$,
indicating a transition to the stripe order.
By contrast, in the FM Kitaev model at the same $J_3$, in addition to these two modes, two further modes at $\Gamma$ and $\Gamma'$ also close their
gaps simultaneously, with the $\Gamma$ mode carrying the dominant spectral weight,
reflecting competing FM and stripe tendencies.
Based on the energetics, the system may select a single-$\bQ$ order (FM or stripe) or realize a multi-$\bQ$ state combining both.
We therefore refer to this regime as FM+stripe.
The FM+stripe regime can also be understood via the $\scT_4$ duality transformation:
applying $\scT_4$ to a stripe order with wave vector $\M_2'$ (see \cref{fig:schematic}(c)), 
whose spins lie in the $x$-$y$ plane (with the spin $z$ component suppressed for energetic reasons),
yields a state in which the spin $x$ components align ferromagnetically while 
the spin $y$ components exhibit a stripe modulation.
Depending on whether the original stripe order has one or two finite spin components,
the $\scT_4$-mapped phase corresponds to a single-$\bQ$ order (FM or stripe) or a multi-$\bQ$ state with coexisting FM and stripe orders 
(see also \Appref{sec:T_4}).


Therefore, our mean-field plus RPA calculations of the spin DSF for KSL reveal both the high-energy Majorana continuum and
low-energy paramagnon-like collective modes, whose condensation reliably characterizes the system's tendency toward magnetic ordering.
Remarkably, at the same critical value of $J_3$, a pair of $\scT_4$-related magnetic orders emerge in the FM and AFM Kitaev models,
in full agreement with the exact $\scT_4$ duality of the model.
While the precise nature (single-$\bQ$ or multi-$\bQ$) of the AFM+zigzag and FM+stripe regimes cannot be resolved
from our RPA calculations, preliminary variational Monte Carlo results suggest that single-$\bQ$ states are favored~\cite{Wang-prep}.

\begin{figure}
\centering
\includegraphics[width=0.5 \textwidth]{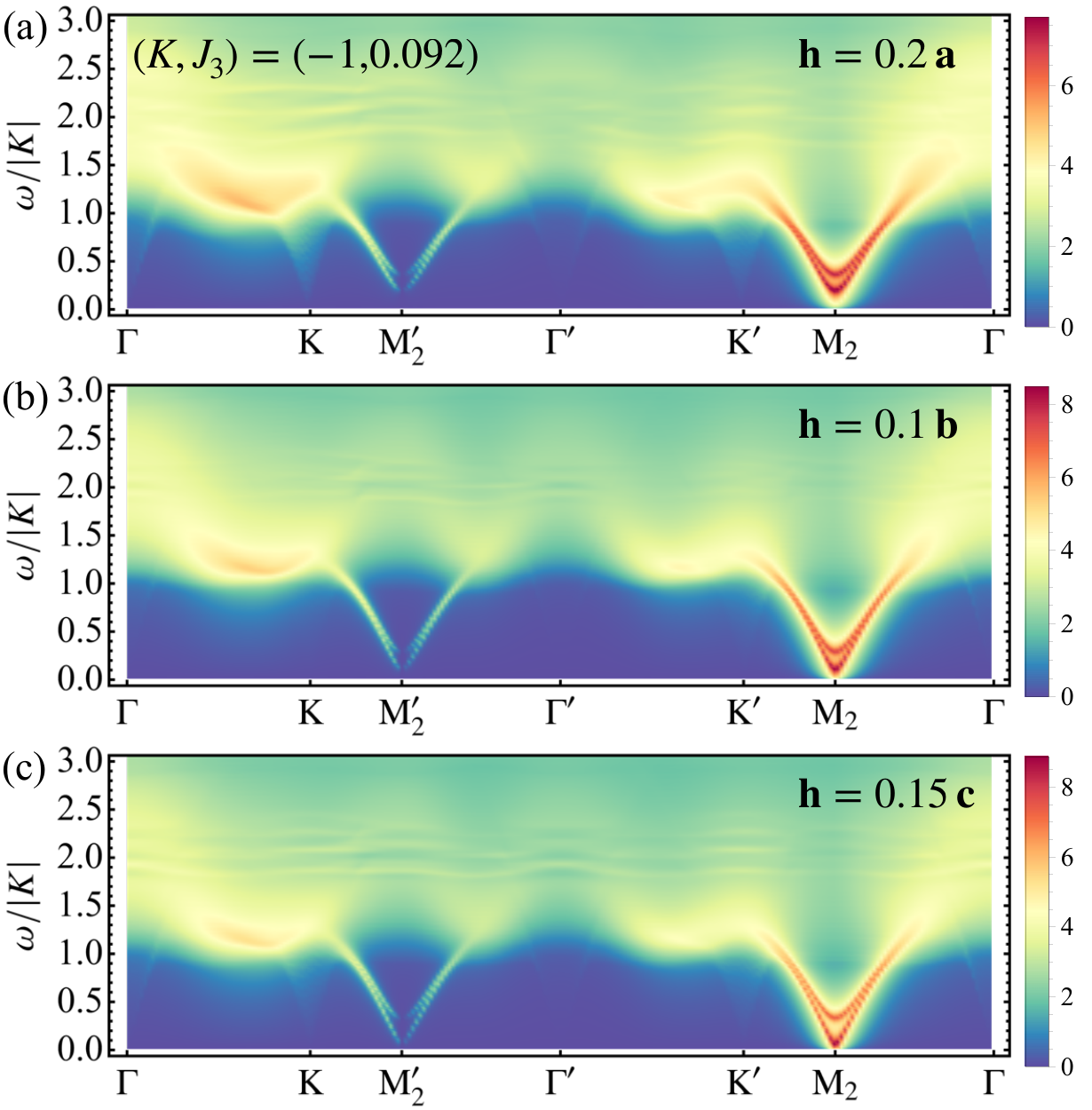}
\caption{Spin DSF for $(K,J_3) = (-1,0.092)$ in the presence of a finite external magnetic field. 
For fields applied along each of the three crystallographic directions, the $\M$ mode softens under the magnetic field,
signaling a field-induced transition to the zigzag order.
}\label{fig:S-h-zz}
\end{figure}

\begin{figure}
\centering
\includegraphics[width=0.5 \textwidth]{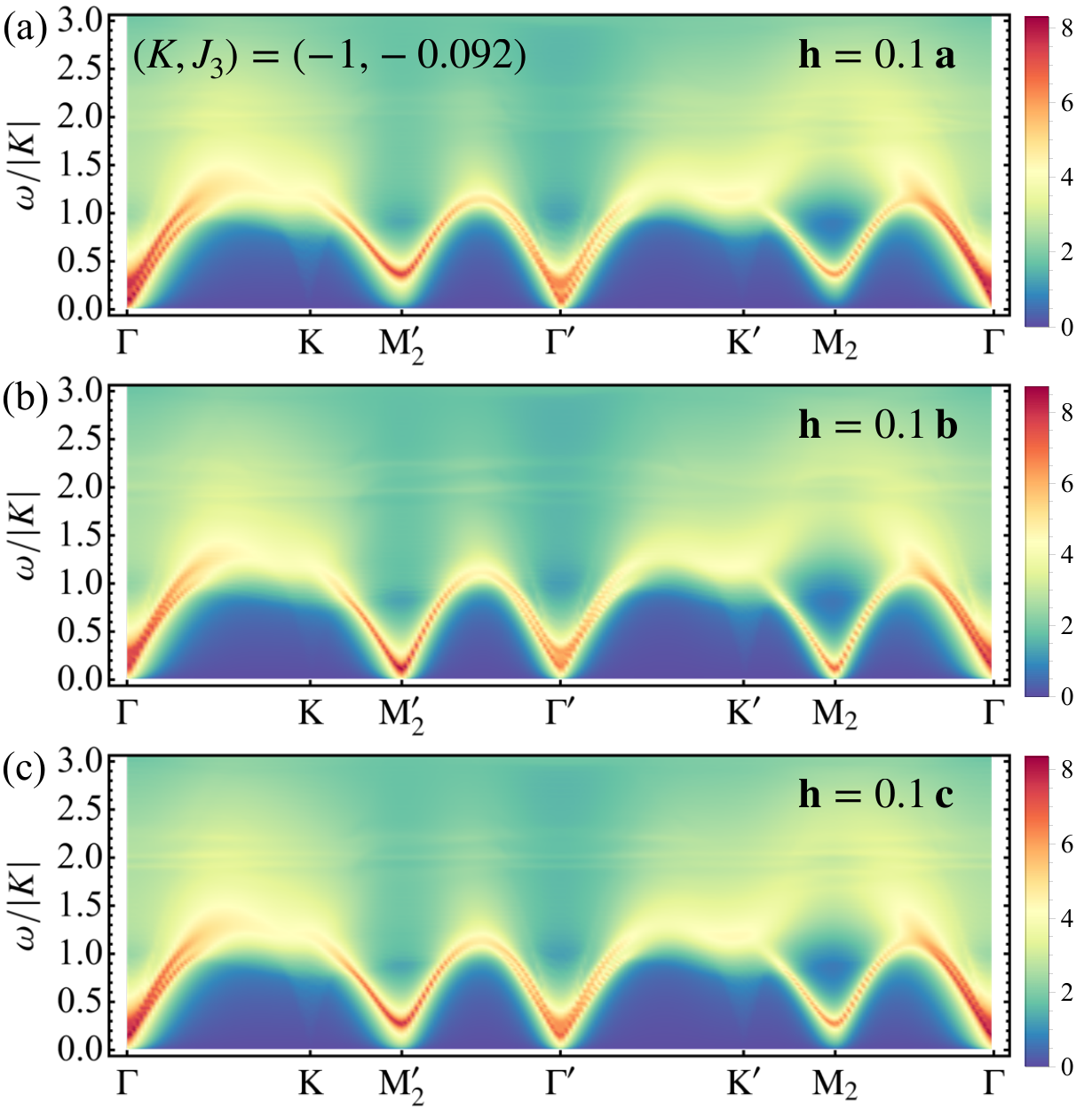}
\caption{Spin DSF for $(K,J_3) = (-1,-0.092)$ in the presence of a finite external magnetic field. 
For fields applied along both (a) $\ba$ and (c) $\bc$ directions, the $\Gamma$ and $\Gamma'$ modes soften
while the $\M$ and $\M'$ modes get lifted, indicating a transition into the FM order.
In contrast, for (b) $\bh \parallel \bb$, all modes get softened, signaling a field-induced transition to the 
FM+stripe regime.
}\label{fig:S-h-FM-stripe}
\end{figure}

\subsection{Magnetic-field effects}
We have also examined the effect of magnetic fields on the $J_3$-perturbed KSL by computing the spin DSF 
in the presence of an external field applied along each of the three crystallographic directions.
It was found that the field generally destabilizes the KSL and enhances the tendency toward magnetic ordering
by softening some or all of the existing low-energy collective modes.

\cref{fig:S-h-zz} shows the impact of a magnetic field on the spin DSF in the FM Kitaev model at $J_3 = 0.092$, 
just below the critical value for the zigzag transition. 
For fields applied along any of the three crystallographic directions, the gap of the $\M$ mode is reduced,
signaling a field-induced tendency toward the zigzag order.
On the opposite side of the $J_3$ axis, at $J_3 = -0.092$, before entering the FM+stripe phase,
the effect of the field depends strongly on its direction (see \cref{fig:S-h-FM-stripe}).
A field along the $\bb$ direction softens all collective modes at $\Gamma$, $\Gamma'$, $\M'$, and $\M$, promoting the
transition to the FM+stripe phase. In contrast, fields along the $\ba$ or $\bc$ directions
only soften the $\Gamma$ and $\Gamma'$ modes while lifting the $\M'$ and $\M$ modes, favoring an FM order.

\begin{figure}
\centering
\includegraphics[width=0.5 \textwidth]{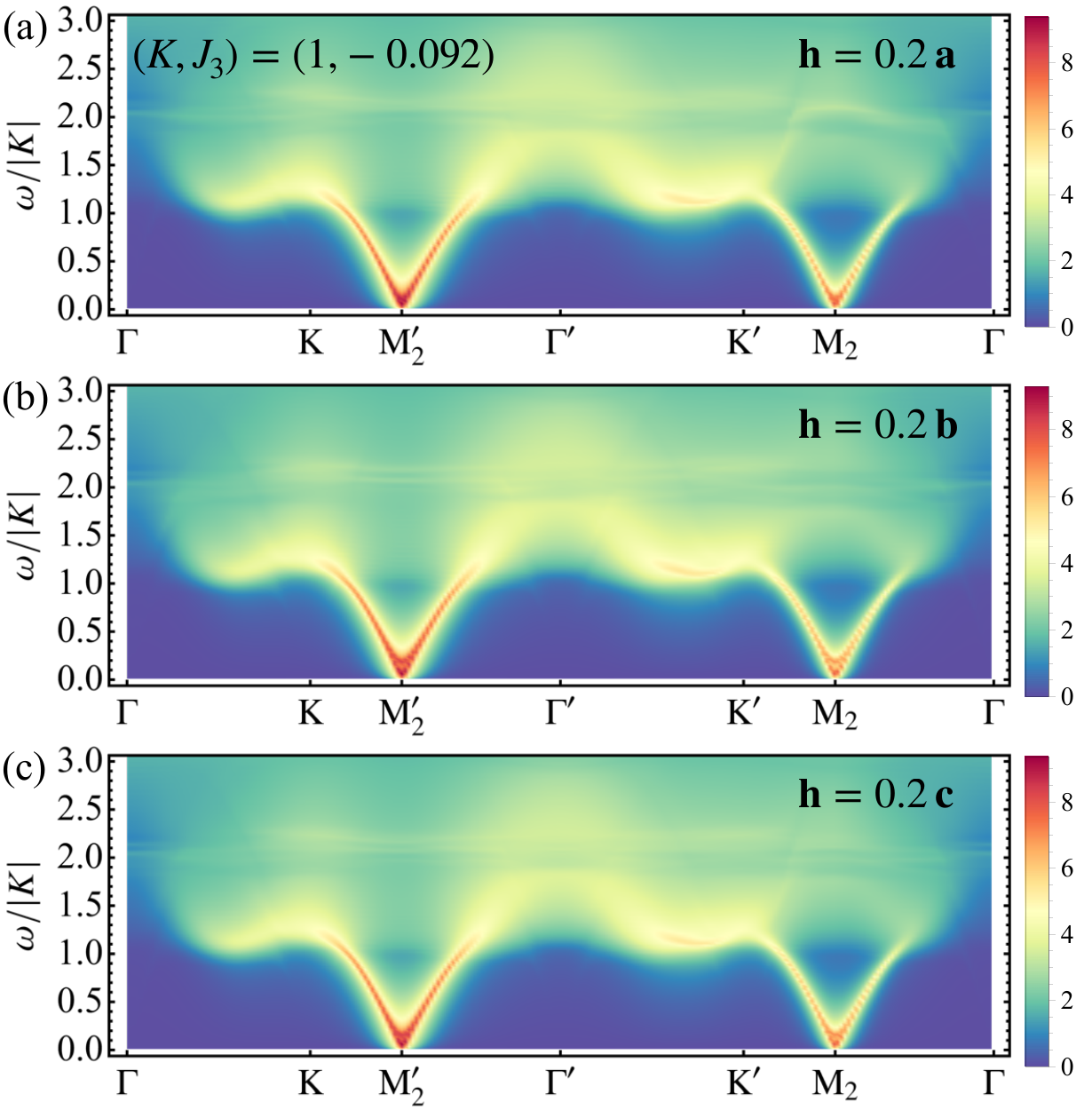}
\caption{Spin DSF for $(K,J_3) = (1, -0.092)$ in the presence of a finite external magnetic field. 
For fields applied along each of the three crystallographic directions, both the $\M'$ and $\M$ modes soften under the magnetic field,
signaling a field-induced transition to the stripe order.
}\label{fig:S-h-stripe}
\end{figure}

\begin{figure}
\centering
\includegraphics[width=0.5 \textwidth]{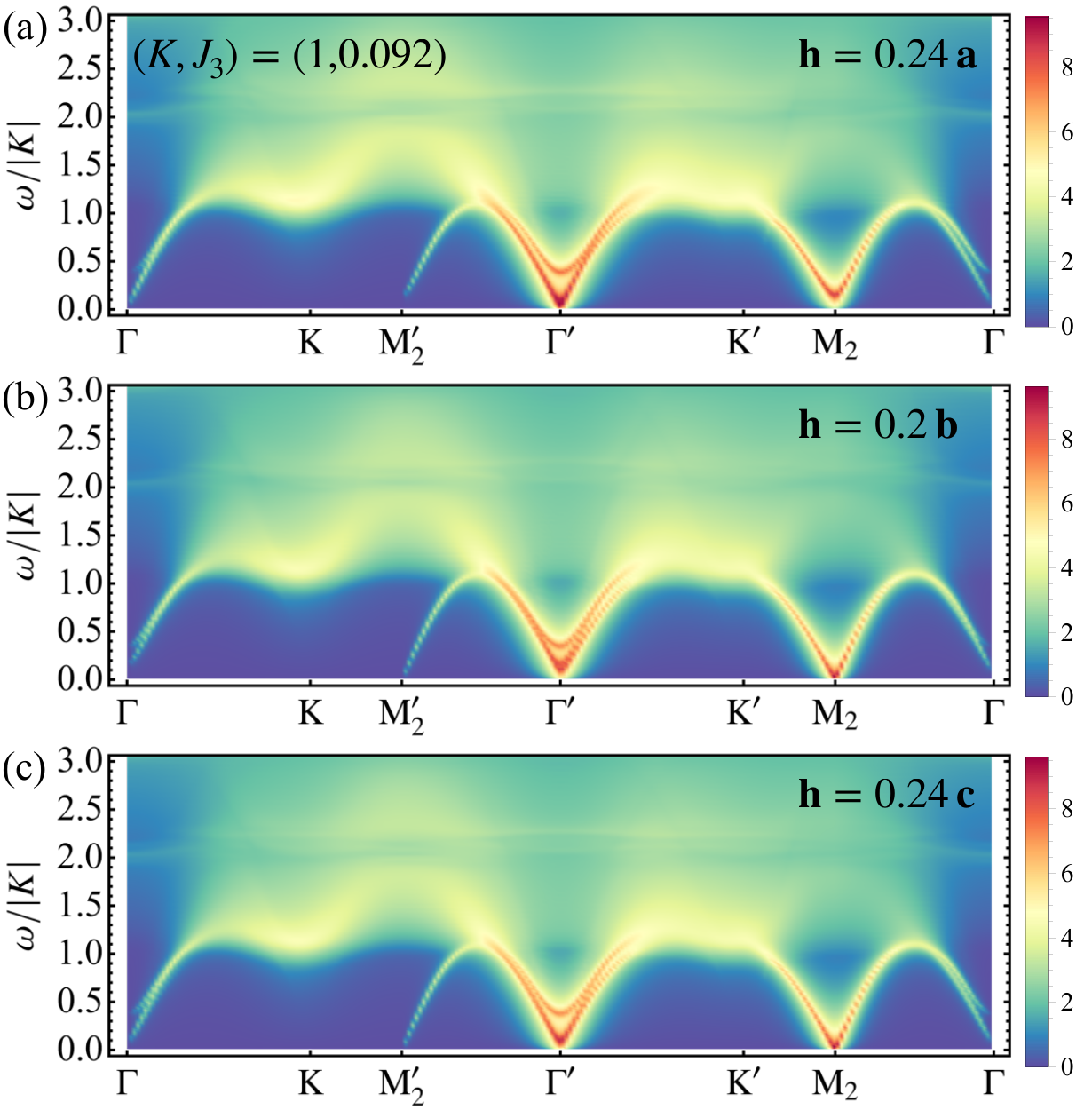}
\caption{Spin DSF for $(K,J_3)=(1,0.092)$ in a finite magnetic field. 
Both the $\Gamma'$ and $\M$ modes soften, but the order in which their gaps close depends on the field direction. 
For a field along (a) $\ba$, the $\Gamma'$ mode condenses first, indicating a transition to a N\'eel AFM phase.
For fields along (b) $\bb$ and (c) $\bc$, the $\M$ gap closes first, signaling a transition to the zigzag phase.
}\label{fig:S-h-AFM-zz}
\end{figure}

Similar field effects are observed in the AFM Kitaev model. \cref{fig:S-h-stripe} shows the spin DSF at $J_3 = -0.092$,
just before the transition to the stripe phase. For fields applied along any crystallographic direction,
the gaps at both the $\M'$ and $\M$ points vanish at $|\bh| \approx 0.2$, indicating a field-induced transition to the stripe order.
At $J_3 = 0.092$, below the critical value for the AFM+zigzag transition,
both the $\Gamma'$ and $\M$ modes are softened by a magnetic field, but the order in which their gaps close
depends on the field direction, as shown in \cref{fig:S-h-AFM-zz}.
For a field along $\ba$, the gap of the $\Gamma'$ mode closes first at $\bh = 0.24\, \ba$, signaling a transition to a N\'eel AFM phase.
In contrast, for fields along $\bb$ and $\bc$, the $\M$ mode condenses first, indicating a transition to the zigzag order.

\section{Raman scattering signatures}\label{sec:Raman}
Raman scattering provides a complementary dynamical probe of QSLs.
In general, QSLs are expected to exhibit only a weak dependence of the Raman intensity on the polarization angle,
in contrast to magnetically ordered states~\cite{Devereaux2007-md,Cepas2008-sa}.
More importantly, fractionalized excitations can also manifest themselves in the Raman response.
For example, in the pure Kitaev model the Raman operator excites only the itinerant Majorana fermions, making
Raman scattering an ideal probe of their fermionic nature~\cite{Nasu2016-sk}.
In this section, we analyze the Raman response of the KSL under a weak $J_3$ perturbation.

\subsection{Raman operator}
We begin by introducing the Raman operator for the $K$–$J_3$ model.
Within the Loudon-Fleury approximation~\cite{Fleury1968-yc}, the Raman operator for the $K$-$J_3$ model contains
two parts, $\scR = \scR_K + \scR_{J_3}$, where
\begin{subequations}
    \begin{align}
        & \scR_K = \sum_{\langle i,j \rangle_\alpha} \lambda_K \, ( \bepsilon_{\text{in}} \cdot \bd_{i,j} ) ( \bepsilon_{\text{out}} \cdot \bd_{i,j} )
        \, \sigma_{i}^\alpha \sigma_j^\alpha, \label{eq:R_K} \\
        & \scR_{J_3} = \sum_{\langle i,j \rangle_3} \lambda_{J_3}\, ( \bepsilon_{\text{in}} \cdot \bd_{i,j} ) ( \bepsilon_{\text{out}} \cdot \bd_{i,j} )
        \, \bm{\sigma}_i \cdot \bm{\sigma}_j. \label{eq:R_J_3}
    \end{align}
\end{subequations}
Here $\bepsilon_{\text{in}}$ ($\bepsilon_{\text{out}}$) denotes the polarization vector of the incoming (outgoing) light, 
which is taken to lie within the honeycomb plane.
$\bd_{i,j}$ is the relative position vector between sites $i$ and $j$.
The constants $\lambda_K$ and $\lambda_{J_3}$ are proportional to $K$ and $J_3$, respectively.
Their ratio $g \equiv \lambda_{J_3}/\lambda_{K} = J_3/K$ therefore
serves as a small parameter in the $J_3$-perturbed KSL.
For simplicity, we factor out the overall coupling constant $\lambda_K$, equivalent to setting $\lambda_K = 1$ and $\lambda_{J_3} = g$.
To ensure the validity of our perturbative analysis, which assumes the KSL remains stable under the small $J_3$ interaction,
we choose $g = 0.05$, safely below the critical value $|J_3/K| = 0.094$ for the transition into ordered phases 
determined from our mean-field plus RPA calculation.


The two contributions to the Raman operator generate different types of excitations in the KSL.
While the Kitaev term ($\scR_K$) excites only the itinerant Majorana fermions $\gamma_0$, the $J_3$ term
($\scR_{J_3}$) inevitably creates vison excitations, i.e., excites the gauge Majoranas $\gamma_{x,y,z}$.
As a consequence, their Raman responses display qualitatively different features, which will be analyzed in the following.

\begin{figure}[t]
\centering
\includegraphics[width=0.45 \textwidth]{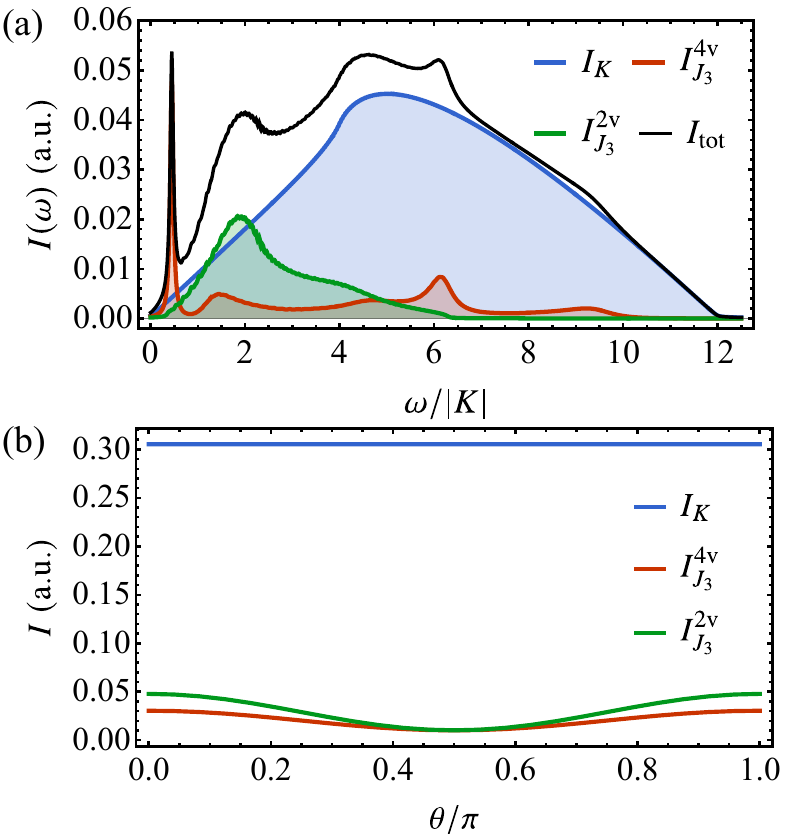}
\caption{(a) Raman response from $\scR_K$ ($I_K$) and $\scR_{J_3}$.
The latter can be further decomposed into contributions from two-vison ($I^\twov_{J_3}$) and four-vison ($I^\fourv_{J_3}$) excitations.
Here, we take $\bepsilon_\text{in} = \bepsilon_\text{out} \parallel \bb$ and $g=0.05$.
(b) Total intensity as a function of the angle between $\bepsilon_\text{in}$ and $\bepsilon_\text{out}$,
with $\bepsilon_\text{in} \parallel \bb$.
}\label{fig:Raman}
\end{figure}

\subsection{Raman signals}
The Raman response of the model is defined as
\begin{equation}
    I(\omega) = \int dt\, e^{i \omega t} \langle \scR(t) \scR(0) \rangle.
\end{equation}
At zero temperature, the expectation value is evaluated with respect to the ground state $|\Phi_0 \rangle$.
Since our focus is the small-$J_3$ regime, where the ground state $| \Phi_0 \rangle$ is adiabatically connected to that of the
pure Kitaev model $| \Omega \rangle$, the correlator
$\langle \scR(t) \scR(0) \rangle$ can be evaluated by treating the $J_3$ interaction ($H_{J_3}$) as a perturbation to
the Kitaev Hamiltonian ($H_K$)~\cite{Knolle2014-ma}. In the interaction picture, it can be written as
\begin{align}
    \langle \Omega | U_I(-\infty,t) \scR^{(0)}(t) U_I(t,0) \scR^{(0)}(0) U_I(0,-\infty) |\Omega \rangle.
\end{align}
Here $\scR^{(0)}(t) = e^{i H_K t}\, \scR \, e^{-i H_K t}$, and the time-evolution operator is
\begin{align}
    U_I(t,t') & = T \, \exp(-i \int_{t'}^t dt_1 \, H_{J_3}^{(0)}(t_1) ) \nonumber \\
    & = 1 - i \int_{t'}^t dt_1 \, H_{J_3}^{(0)}(t_1) + \dots
\end{align}
To leading order in $g$, and retaining only terms involving the lowest-order multi-$\gamma_0$ correlation functions~\footnote{The
multi-$\gamma_0$ correlation function scales with a higher power of the density of states of the matter fermions.
Consequently, terms involving higher-order $\gamma_0$ correlators mainly contribute to a broad continuum and are
strongly suppressed at low energies~\cite{Knolle2014-ma}},
the Raman response is dominated by two contributions $I(\omega) = I_K(\omega) + I_{J_3}(\omega)$, with
\begin{subequations}
\begin{align}
    & I_K(\omega) = \int dt\, e^{i \omega t} \langle \Omega | \scR_K^{(0)}(t) \scR_K^{(0)}(0) |\Omega \rangle, \label{eq:I_K} \\
    & I_{J_3}(\omega) = \int dt\, e^{i \omega t} \langle \Omega | \scR_{J_3}^{(0)}(t) \scR_{J_3}^{(0)}(0) |\Omega \rangle. \label{eq:I_J_3}
\end{align}
\end{subequations}

Since the correlators above involve time evolution governed solely by the Kitaev Hamiltonian, they can be evaluated using
the exact solutions of $H_K$. In the Majorana representation,
\begin{align}
    H_K = \sum_{\langle i,j \rangle_\alpha} K \, \hat{u}_{i,j} \, i \gamma_{i,0} \gamma_{j,0},
\end{align}
where for a bond $\langle i,j \rangle_\alpha$ with $i \in A$, we define a complex gauge fermion $\chi_{i,\alpha} \equiv ( \gamma_{i,\alpha}+i\, \gamma_{j,\alpha})$,
whose parity corresponds to the $\bbZ_2$ gauge field $\hat{u}_{i,j} \equiv -i \gamma_{i,\alpha} \gamma_{j,\alpha} = 1-2 \chi_{i,\alpha}^\dagger \chi_{i,\alpha}$.
We also introduce a complex matter fermion within each unit cell, $a_{\br} = (\gamma_{\br,A,0} + i\, \gamma_{\br,B,0})$
(see \cref{fig:schematic}(a)).
Since $[ H_K, \hat{u}_{i,j} ] = 0$, $H_K$ can be diagonalized within each fixed gauge-field sector $\{ u_{i,j} \}$, where it reduces to a matter-fermion
Bogoliubov-de Gennes (BdG) Hamiltonian
\begin{equation}
    H^a[\{ u_{i,j }\}] \equiv  \sum_{\langle i,j \rangle_\alpha} K \, u_{i,j} \, i \gamma_{i,0} \gamma_{j,0}.
\end{equation}
The ground state of $H_K$ lies in the vacuum sector of $\chi$ fermions, satisfying $\hat{u}_{i,j}|0_\chi \rangle = |0_\chi \rangle$.
In this sector, the matter-fermion BdG Hamiltonian takes the form
\begin{align}
    H^a[\{  1 \}] & = \frac{1}{2} \sum_{\bk} ( a_\bk^\dagger, a_{-\bk} )
    \begin{pmatrix}
        \Re f_{\bk} & -i \, \Im f_{\bk} \\
        i \, \Im f_{\bk} & - \Re f_{\bk}
    \end{pmatrix}
    \begin{pmatrix}
        a_{\bk} \\
        a_{-\bk}^\dagger
    \end{pmatrix} \nonumber \\
    & = \sum_{\bk} \epsilon_{\bk} (\alpha_{\bk}^\dagger \alpha_{\bk} - \frac{1}{2}),
\end{align}
where $f_{\bk} = 2K (1 + e^{i \bk \cdot \ba_1} + e^{i \bk \cdot \ba_2} )$, and $\epsilon_\bk = |f_\bk|$ is the energy of the Bogoliubov quasiparticle $\alpha_{\bk}$.
Its ground state $| \Psi^a_{0} \rangle$ satisfies $\alpha_\bk |\Psi^a_{0} \rangle = 0$.
The ground state of $H_K$ is then $| \Omega \rangle \propto P |0_{\chi}; \Psi^a_{0} \rangle$,
where the projector $P = \prod_j \frac{1}{2} (1 + \gamma_{j,x} \gamma_{j,y} \gamma_{j,z} \gamma_{j,0})$ enforces the physical constraint
$\gamma_{j,x} \gamma_{j,y} \gamma_{j,z} \gamma_{j,0} = 1$.

Since $\scR_K$ does not excite gauge fluxes, the correlator in \cref{eq:I_K} reduces to a matter-fermion correlation function,
\begin{align}
    & \langle \Psi^a_0 | e^{i t H^a_0} \sum_{\langle i,j \rangle_\alpha} (\bepsilon_\text{in} \cdot \bd_{i,j})(\bepsilon_\text{out} \cdot \bd_{i,j})
    \, i \gamma_{i,0}\gamma_{j,0}  \nonumber \\
    & e^{-i t H^a_0} \sum_{\langle k,l \rangle_\beta} (\bepsilon_\text{in} \cdot \bd_{k,l})(\bepsilon_\text{out} \cdot \bd_{k,l})\, i \gamma_{k,0}\gamma_{l,0} | \Psi^a_0 \rangle.
\end{align}
Here we denote $H^a_0 \equiv H^a[\{1\}]$ for brevity.
Evaluating this correlator yields~\cite{Knolle2014-ma}
\begin{align}\label{eq:I_K-result}
    I_K(\omega) = \,4\pi \sum_{\bk} \delta(\omega-2 E_{\bk})\, \frac{\Im[ f_\bk \, v_\bk^*]^2}{E_\bk^2},
\end{align}
with
$v_\bk = (\bepsilon_\text{in} \cdot \bdelta_x)(\bepsilon_\text{out} \cdot \bdelta_x) 
+ (\bepsilon_\text{in} \cdot \bdelta_y)(\bepsilon_\text{out} \cdot \bdelta_y) \, e^{i \bk \cdot \ba_1}
+ (\bepsilon_\text{in} \cdot \bdelta_z)(\bepsilon_\text{out} \cdot \bdelta_z) \, e^{i \bk \cdot \ba_2}$.

\cref{eq:I_K-result} suggests that the overall behavior of $I_K(\omega)$ reflects the two-particle density of states of the matter fermions.
As shown in \cref{fig:Raman}(a), it increases linearly from $\omega=0$, forms a broad continuum peaked around $\omega = 4|K|$,
and terminates at $\omega = 12|K|$.
\cref{fig:Raman}(b) shows the total intensity versus polarization angle, which is independent of polarization, as expected.

\begin{figure}[t]
\centering
\includegraphics[width=0.45 \textwidth]{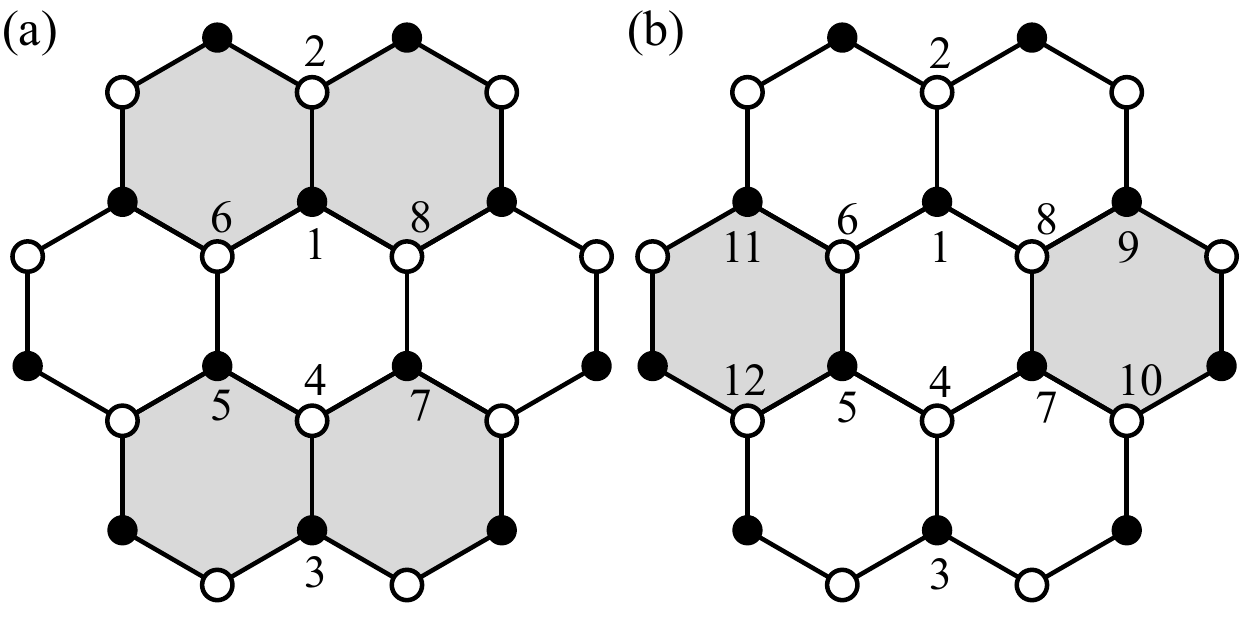}
\caption{Examples of intermediate vison configurations (gray hexagons) generated by $\scR_{J_3}$.
(a) Four-vison configuration generated by $\sigma_1^z \sigma_4^z$.
(b) Two-vison configuration generated by $\sigma_5^z \sigma_8^z$ and $\sigma_7^z \sigma_6^z$.
}\label{fig:Raman-vison}
\end{figure}

Unlike $\scR_K$, $\scR_{J_3}$ creates vison excitations.
Depending on the orientation of $\langle i,j \rangle_3$ and the spin components of the bilinear terms in \cref{eq:R_J_3},
either two- or four-vison states are generated.
In the former case the visons reside on next-NN plaquettes, while in the latter they form two NN vison pairs (see \cref{fig:Raman-vison}).
Accordingly, the $\scR_{J_3}$ correlator in \cref{eq:I_J_3} can be decomposed as
\begin{align}
    \langle \Omega | \scR_{J_3}^{(0)}(t) \scR_{J_3}^{(0)}(0) | \Omega \rangle = g^2 \sum_{\br \in A,\alpha} F^{\twov}_{\br,\alpha}(t)
    + F^{\fourv}_{\br,\alpha}(t).
\end{align}

In $F^\twov_{\br,\alpha}(t)$, the two next-NN visons are separated by two $\alpha$ bonds emanating from $\br$
and $\br+ \varepsilon_{\alpha \beta \gamma} (\bdelta_{\beta} - \bdelta_{\gamma})/2$, respectively.
In $F^\fourv_{\br,\alpha}(t)$, the two NN vison pairs reside on the $\alpha$ bonds emanating from $\br$
and $\br - 3 \bdelta_\alpha$, respectively.
\cref{fig:Raman-vison} shows typical vison configurations associated with $F^{\twov}_{\br,z}$ and $F^{\fourv}_{\br,z}$,
whose expression are derived below.
The results for other orientations of visons can be obtained similarly.

The four visons in $F^\fourv_{1,z}(t)$ (c.f. \cref{fig:Raman-vison}(a)) are created by $\sigma_1^z \sigma_4^z$.
Therefore
\begin{subequations}
\begin{align}
    F^\fourv_{1,z}(t) & = (2\bdelta_z \cdot \bepsilon_\text{in})^2(2\bdelta_z \cdot \bepsilon_\text{out})^2 \, C^\fourv_{1,z}(t), \\
    C^\fourv_{1,z}(t) & = \langle \Omega| e^{i H_K t} \sigma_1^z \sigma_4^z e^{-i H_K t} \sigma_1^z \sigma_4^z | \Omega \rangle.
\end{align}
\end{subequations}
Utilizing the Majorana fermion representation, the correlator can be written as
\begin{align}
     C^\fourv_{1,z}(t) = \langle \Psi^a_0 | e^{i H^a_0 t} e^{-i [H^a_0 + V^\fourv] t} | \Psi^a_{0} \rangle. \label{eq:C^4v}
\end{align}
Here the local perturbation $V^\fourv$, corresponding to four additional visons, is given by
\begin{align}
    V^\fourv = -2K \, [ i \gamma_{1,0} ( \gamma_{6,0} + \gamma_{8,0} )
    + i ( \gamma_{5,0} + \gamma_{7,0} ) \gamma_{4,0} ].
\end{align}
The problem thus reduces to that of a local quantum quench, which can be evaluated by inserting a complete set of eigenbasis
$\{ | \psi \rangle \}$ of $H^a_0 + V^\fourv$. The Fourier transform of $C^\fourv_{1,z}$ is then
\begin{align}
    C^\fourv_{1,z}(\omega) = \sum_{\psi} 2\pi \, \delta(\omega-[E_\psi - E_0]) \, |\langle \Psi^a_0| \psi \rangle|^2. \label{eq:C^4v-Lehmann}
\end{align}

Since the ground state $| \Psi^{a}_{0,\fourv} \rangle$ of $H^a_0 + V^\fourv$ has the same fermion parity as $| \Psi^a_0 \rangle$,
states with an even number of Bogoliubov quasiparticle of $H^a_0 + V^\fourv$
(with annihilation operators $\alpha_{i,\fourv}$, $i=1,\dots,N$) contribute to the sum over $\psi$ in \cref{eq:C^4v-Lehmann}.
However, because the fermionic density of states vanishes at low energies, contributions from states with many quasiparticles 
are strongly suppressed.
We therefore retain only the leading contributions from the (zero-quasiparticle) ground state $|\Psi^{a}_{0,\fourv} \rangle$ and the two-quasiparticle states
$\alpha_{i,\fourv}^\dagger \alpha_{j,\fourv}^\dagger | \Psi^{a}_{0,\fourv} \rangle$~\footnote{In calculating the spin DSF of the Kitaev model, 
this type of approximation has been shown to reproduce the exact results accurately~\cite{Knolle2014-ma,Knolle2015-lx}.}:
\begin{align} \label{eq:C4v}
    C^\fourv_{1,z}(\omega) =& 2\pi \, \delta(\omega - \Delta^\fourv ) \, |\langle \Psi^a_0| \Psi^a_{0,\fourv} \rangle|^2 \nonumber \\
    & + \sum_{i \neq j} 2\pi \, \delta(\omega-[\Delta^\fourv + \epsilon^{\fourv}_{i} + \epsilon^\fourv_{j} ])\, \nonumber \\
    & \times |\langle \Psi^a_0|\alpha_{i,\fourv}^\dagger \alpha_{j,\fourv}^\dagger|\Psi^a_{0,\fourv} \rangle|^2.
\end{align}
Here, $\Delta^\fourv \equiv E^\fourv_0 - E_0 \approx 0.44\, |K|$ is the four-vison excitation energy, 
and $\epsilon^\fourv_i$ denotes the energy of quasiparticle $\alpha_{i,\fourv}$.
The overlaps $\langle \Psi^a_0| \Psi^a_{0,\fourv} \rangle$ and 
$\langle \Psi^a_0|\alpha_{i,\fourv}^\dagger \alpha_{j,\fourv}^\dagger|\Psi^a_{0,\fourv} \rangle$ are given in \Appref{sec:Raman-details}.

\cref{fig:Raman}(a) presents the Raman response $I^\fourv_{J_3}(\omega)$ from $g^2 \sum_{\br,\alpha} F^\fourv_{\br,\alpha}$.
The ground state $|\Psi^a_{0,\fourv} \rangle$ produces a sharp peak at $\Delta^\fourv$,
while the two-quasiparticle states form a continuum with broad maxima near $\omega \approx 1.5 \,|K|$
and $\omega \approx 6 \, |K|$.
Interestingly, a similar Raman response---featuring both a sharp peak and a two-quasiparticle continuum---has been reported for the 
NN Heisenberg interaction~\cite{Knolle2014-ma}, which also excites four visons, albeit with a different spatial distribution.
However, unlike the NN Heisenberg case, the $J_3$ interaction can additionally generate two-vison excitations, which we discuss next.

In $F^\twov_{7,z}$, two visons can be created either by 
$\sigma_5^z \sigma_8^z$ and $\sigma_7^z \sigma_6^z$ (c.f. \cref{fig:Raman-vison}(b)).
This leads to
\begin{align}
    F^\twov_{7,z}(t) = & \sum_{\alpha,\beta = x,y} (2\bdelta_\alpha \cdot \bepsilon_\text{in}) (2\bdelta_\alpha \cdot \bepsilon_\text{out}) 
     \, C^\twov_{7,z;\alpha \beta}(t) \nonumber \\
    & \times (2\bdelta_\beta \cdot \bepsilon_\text{in}) (2\bdelta_\beta \cdot \bepsilon_\text{out}),
\end{align}
with the correlators defined as
\begin{subequations}
\begin{align}
    C^\twov_{7,z;xx}(t) & = \langle \Omega | e^{i H_K t} \sigma_5^z \sigma_8^z e^{-i H_K t} \sigma_5^z \sigma_8^z | \Omega \rangle, \\
    C^\twov_{7,z;xy}(t) & = \langle \Omega | e^{i H_K t} \sigma_5^z \sigma_8^z e^{-i H_K t} \sigma_7^z \sigma_6^z | \Omega \rangle, \\
    C^\twov_{7,z;yx}(t) & = \langle \Omega | e^{i H_K t} \sigma_7^z \sigma_6^z e^{-i H_K t} \sigma_5^z \sigma_8^z | \Omega \rangle, \\
    C^\twov_{7,z;yy}(t) & = \langle \Omega | e^{i H_K t} \sigma_7^z \sigma_6^z e^{-i H_K t} \sigma_7^z \sigma_6^z | \Omega \rangle.
\end{align}
\end{subequations}
Under the Majorana representation, they can be written as
\begin{subequations}
\begin{align}
    C^\twov_{7,z;xx}(t) = & e^{i E_0 t} \langle \Psi^a_0 |  e^{-i [ H^a_0 + V^{\twov}_1 ] t} | \Psi^a_0 \rangle, \label{eq:C_z-xx} \\
    C^\twov_{7,z;xy}(t) = & -e^{i E_0 t} \langle \Psi^a_0 | e^{-i [ H^a_0 + V^{\twov}_1 ] t} \mathcal{O}_{7,z}| \Psi^a_0 \rangle, \label{eq:C_z-xy} \\
    C^\twov_{7,z;yx}(t) = & -e^{i E_0 t} \langle \Psi^a_0 | \mathcal{O}_{7,z} e^{-i [ H^a_0 + V^{\twov}_1 ] t} | \Psi^a_0 \rangle, \label{eq:C_z-yx} \\
    C^\twov_{7,z;yy}(t) = & e^{i E_0 t} \langle \Psi^a_0 |  e^{-i [ H^a_0 + V^{\twov}_2 ] t} | \Psi^a_0 \rangle, \label{eq:C_z-yy}
\end{align}
\end{subequations}
where $\mathcal{O}_{7,z} \equiv ( i \, \gamma_{5,0} \gamma_{8,0} )( i \, \gamma_{7,0} \gamma_{6,0} )$.
Both \(V^{\twov}_1\) and \(V^{\twov}_2\) correspond to the same two-vison configuration, but are different by a gauge transformation:
\begin{subequations}
    \begin{align}
        V^{\twov}_1 & = -2K [ i \gamma_{5,0}( \gamma_{12,0} + \gamma_{4,0} ) + i ( \gamma_{1,0} + \gamma_{9,0} )\gamma_{8,0} ], \\
        V^{\twov}_2 & = -2K [ i \gamma_{7,0}( \gamma_{4,0} + \gamma_{10,0} ) + i ( \gamma_{11,0} + \gamma_{1,0} )\gamma_{6,0} ].
    \end{align}
\end{subequations}
The correlation functions in \cref{eq:C_z-xx,eq:C_z-xy,eq:C_z-yx,eq:C_z-yy} can also be evaluated by inserting the complete eigenbasis of 
$H^a_0 + V^{\twov}_1$ or $H^a_0 + V^{\twov}_2$. 
However, since their ground states $|\Psi^a_{0,\twov,1} \rangle$ and $| \Psi^a_{0,\twov,2} \rangle$ have opposite fermion parity
relative to $| \Psi^a_0 \rangle$, the leading contributions come from single-quasiparticle states 
$\alpha_{i,\twov,n}^\dagger |\Psi^a_{0,\twov,n} \rangle$ ($n=1,2$).
Their Fourier transforms are then given by
\begin{subequations}
\begin{align}
    C^\twov_{7,z;xx}(\omega) = & 2\pi \sum_i \delta(\omega - \Delta^{\twov} - \epsilon^{\twov}_i ) 
    |\langle \Psi^a_0 | \alpha_{i,\twov,1}^\dagger | \Psi^a_{0,\twov,1} \rangle|^2, \label{eq:C2v_xx} \\
    C^\twov_{7,z;xy}(\omega) = & -2\pi \sum_i \delta(\omega - \Delta^{\twov} - \epsilon^{\twov}_i ) \nonumber \\
    & \langle  \Psi^a_0 | \alpha_{i,\twov,1}^\dagger | \Psi^a_{0,\twov,1} \rangle
    \langle \Psi^a_{0,\twov,1} | \alpha_{i,\twov,1} \mathcal{O}_{7,z} | \Psi^a_0 \rangle, \label{eq:C2v_xy} \\
    C^\twov_{7,z;yz}(\omega) = & -2\pi \sum_i \delta(\omega - \Delta^{\twov} - \epsilon^{\twov}_i ) \nonumber \\
    & \langle \Psi^a_0| \mathcal{O}_{7,z} \alpha_{i,\twov ,1}^\dagger |\Psi^a_{0,\twov ,1} \rangle
    \langle \Psi^a_{0,\twov ,1} | \alpha_{i,\twov ,1} | \Psi^a_0 \rangle, \label{eq:C2v_yx} \\
    C^\twov_{7,z;yy}(\omega) = & 2\pi \sum_i \delta(\omega - \Delta^{\twov} - \epsilon^{\twov}_i ) 
    |\langle \Psi^a_0 | \alpha_{i,\twov,2}^\dagger | \Psi^a_{0,\twov,2} \rangle|^2. \label{eq:C2v_yy}
\end{align}
\end{subequations}
Here, $\Delta^\twov \equiv E^\twov_0 - E_0 \approx 0.30\, |K|$ is the two-vison excitation gap,
and the overlaps appearing above are given in \Appref{sec:Raman-details}.

The Raman response $I^\twov_{J_3}(\omega)$ from $g^2 \sum_{\br,\alpha} F^\twov_{\br,\alpha}$ is shown in \cref{fig:Raman}(a).
Notably, the continuum mirrors the matter-fermion density of states:
it rises linearly above $\Delta^\twov$ and exhibits a broad peak near $\omega \approx  2\,|K|$, close to the van Hove singularity.
This is particularly interesting because a single matter fermion cannot be created by a local probe.
%
Notably, the Raman signal from $\scR_K$ contains only the isotropic $E_g$ channel, whereas that from $\scR_{J_3}$ 
contains both a polarization-dependent $A_{1g}$ channel, proportional to $\cos^2(\theta_\text{in}-\theta_\text{out})$, 
and a polarization-independent $E_g$ channel (see \cref{fig:Raman}(b))~\cite{Cepas2008-sa}.

\section{Discussion}\label{sec:discuss}
In this work, we have investigated the dynamical responses of the KSL in the presence of a third-NN Heisenberg ($J_3$) interaction.
In particular, we calculate its spin DSF and Raman response.

Within a self-consistent parton mean-field plus RPA framework, we find that a hallmark of the $J_3$-perturbed KSL is 
the emergence of coherent paramagnon-like collective modes coexisting with the high-energy Majorana continuum,
even though such modes typically being 
associated with proximate magnetic order.
These paramagnon modes soften with increasing $J_3$ and eventually condense at a critical coupling,
providing a natural explanation for the transition from the KSL to magnetically ordered states.

A central result is that both FM and AFM Kitaev-$J_3$ models develop magnetic order at a common critical value $|J_3| = 0.094|K|$.
Specifically, at $J_3 = 0.094|K|$, the FM Kitaev model transitions to a zigzag phase, 
while in the AFM Kitaev model, the softening of modes at the $\Gamma'$ and $\M$ points signals the emergence of an AFM+zigzag regime,
which may correspond either to a near-degenerate single-$\bQ$ phase (AFM or zigzag) or to a multi-$\bQ$ phase with coexisting AFM
and zigzag orders.
Correspondingly, at $J_3 = -0.094|K|$, the AFM Kitaev model exhibits stripe order,
while the FM Kitaev model enters the FM+stripe regime characterized by competing tendencies toward FM and stripe order.
Remarkably, in each case, the phases realized in the FM and AFM Kitaev models 
are related by the exact $\scT_4$ duality of the model ($K \rightarrow -K$),
confirming that our predictions fully respect this exact symmetry.
Resolving the precise nature of the AFM+zigzag and FM+stripe regimes lies beyond the scope of present framework
and requires further numerical investigation, for example via exact diagonalization or density-matrix renormalization group calculations.
Moreover, external magnetic fields could soften some or all of the low-energy paramagnon modes present at zero field, 
thereby facilitating the transition from the KSL to magnetic order.
The general mean-field plus RPA mechanism is similar to that discussed for KSL perturbed by NN non-Kitaev interactions~\cite{Rao2025-bg}:
non-Kitaev interactions generate low-energy collective modes whose softening signals magnetic ordering. 
In the present $K$-$J_3$ model, however, the exact $\mathcal{T}_4$ duality further constrains these instabilities, 
yielding the same critical $|J_3|$ in the FM and AFM Kitaev models and relating the resulting magnetic ordered states.

Complementary to the spin DSF, the Raman scattering provides another experimental probe of fractionalized excitations. 
A perturbative analysis shows that the Raman intensity receives separate contributions from a Kitaev-like vertex $\scR_K$ and 
a $J_3$-like vertex $\scR_{J_3}$.
While the response from $\scR_K$ originates solely from the itinerant matter Majorana fermions,
the $\scR_{J_3}$ contribution reflects both the matter Majoranas and vison excitations.

Furthermore, the $\scR_{J_3}$ contribution can be decomposed into two components 
associated with two-vison ($I_{J_3}^\twov$) and four-vison ($I_{J_3}^\fourv$) processes.
The four-vison component exhibits a sharp peak at the four-vison excitation gap, accompanied by
a broad two-fermion continuum, similar to the Raman response induced by a NN Heisenberg interaction~\cite{Knolle2014-ma}.
In contrast, the two-vison channel forms a continuum reminiscent of the single-fermion density of states---an intriguing result,
given that a local probe cannot excite an isolated matter fermion.
Meanwhile, the polarization-dependent Raman response offers a direct experimental signature of fractionalized excitations modified by
the $J_3$ interaction, revealing distinct contributions from itinerant Majorana fermions and visons.

Overall, our results provide a unified dynamical picture of the $J_3$-perturbed KSL,
in which both the spin DSF and Raman response reveal the interplay between fractionalized excitations.
The emergence and softening of paramagnon modes capture the tendency of KSL toward magnetic ordering
and yield a phase diagram, while the Raman response further resolves the underlying vison excitations.
We expect that these results offer useful guidance for interpreting dynamical responses in field-induced disordered regimes of candidate Kitaev materials, 
particularly in systems with sizable $J_3$ interactions, such as Na$_2$Co$_2$TeO$_6$.


\acknowledgments
We thank Peng Rao for fruitful discussions.
C.C. acknowledges support from the National Natural Science Foundation of China (Grants No.~12404175 and No.~12247101),
the Fundamental Research Funds for the Central Universities (Grant No,~lzujbky-2025-jdzx07),
the Natural Science Foundation of Gansu Province (No.~22JR5RA389, No.~25JRRA799).
J.W. acknowledges support from the National Natural Science Foundation of China under Grant No.~12404170 and the start-up grant at HZNU.
We thank Beijing Paratera Co., Ltd. for providing HPC resources that contributed to the numerical results reported in this paper.

\appendix
%

\begin{figure*}[htbp]
\centering
\includegraphics[width=0.65 \textwidth]{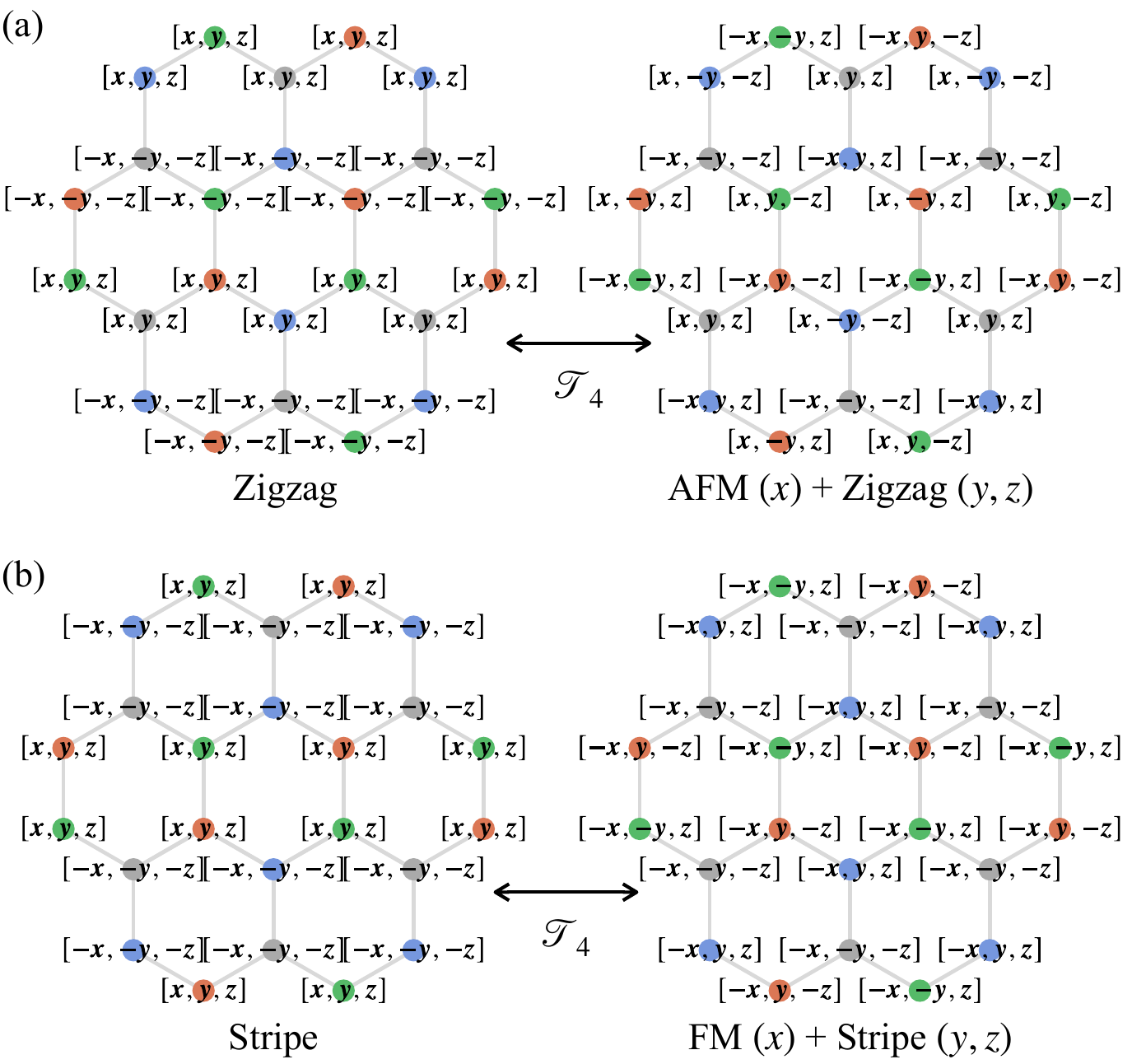}
\caption{Magnetic order patterns induced by the $J_3$ interaction and related by the $\scT_4$ transformation.
The four sublattice involved in $\scT_4$ are indicated by gray (sublattice-1), red (sublattice-2),
green (sublattice-3), and blue (sublattice-4) colors.
(a) Zigzag order with wave vector $\mathrm{M_2}$ and the corresponding phase obtained via the $\scT_4$ transformation.
In the latter, the spin $x$ components form a N\'eel AFM pattern, while the spin $y$ and spin $z$ components
form zigzag patterns with wave vectors $\mathrm{M_1}$ and $\mathrm{M}_3$, respectively (see \cref{fig:schematic}).
(b) stripe order with wave vector $\mathrm{M_2'}$ and the corresponding phase exhibiting both FM and stripe components.
In the latter, the spin $x$ components form a uniform FM pattern, whereas the spin $y$ and spin $z$ components develop
stripe modulations with wave vectors $\mathrm{M_1'} = \mathrm{M_1}+\mathbf{b}_2$ and $\mathrm{M_3'} = \mathrm{M_3}+\mathbf{b}_2$, respectively
(see \cref{fig:schematic}).
For the specific zigzag and stripe patterns considered above, $z$ is expected to vanish from on energetic grounds.
}\label{fig:T_4}
\end{figure*}

\section{Mean-field Hamiltonian in an external magnetic field}\label{sec:MF-with-B}
In this section, we discuss the mean-field Hamiltonian for the KSL in the presence of an external magnetic field, 
which may be applied along any of the three crystallographic directions.
For each case, the symmetries preserved by the model (see \cref{tab:symmetry}) constrain
the number of independent mean-field parameters.
The mean-field Hamiltonian consists of three parts, given in \cref{eq:decouple-Kitaev,eq:decouple-J3,eq:Lagrange}.

1). $\bh \parallel \ba$. In this case, the system preserves the translations $T_{1,2}$ and $\scT \scM_b$ symmetry.
This implies
\begin{align}
    & u_{x,0} = u_{y,0}, \, u_{x,x} = u_{y,y}, \, v_{x,0} = v_{y,0} ,\nonumber \\
    & v_{x,x} = v_{y,y}, \, v_{x,y} = v_{y,x}, \, v_{x,z} = v_{y,z}, \, v_{z,x} = v_{z,y}, \nonumber \\
    & m_x = m_y, \, \lambda_x = \lambda_y.
\end{align}
Thus, 15 independent mean-field parameters must be determined self-consistently.

2). $\bh \parallel \bb$. In this case, the system preserves the translations $T_{1,2}$ and mirror symmetry $\scM_b$,
leading to
\begin{align}
    & u_{x,0} = u_{y,0}, \, u_{x,x} = u_{y,y}, v_{x,0} = v_{y,0}, \nonumber \\
    & v_{x,x} = v_{y,y}, \, v_{x,y} = v_{y,x}, \, v_{x,z} = v_{y,z}, \, v_{z,x} = v_{z,y}, \nonumber \\
    & m_x = -m_y, \, m_z = 0, \, \lambda_x = -\lambda_y, \, \lambda_z = 0.
\end{align}
Accordingly, there are 13 independent mean-field parameters.

3). $\bh \parallel \bc$. In this case, the system preserves the translation $T_{1,2}$, the $C_6$ rotation, and the $\scT \scM_b$ symmetry.
This yields
\begin{align}
    & u_{x,0} = u_{y,0} = u_{z,0}, \, u_{x,x} = u_{y,y} = u_{z,z}, \nonumber \\
    & v_{x,0} = v_{y,0} = v_{z,0}, \, v_{x,x} = v_{y,y} = v_{z,z}, \nonumber \\
    & v_{x,y} = v_{y,z} = v_{z,x} = v_{x,z} = v_{y,x} = v_{z,y}, \nonumber \\
    & m_x = m_y = m_z, \, \lambda_x = \lambda_y = \lambda_z.
\end{align}
In this case, 7 independent mean-field parameters are to be determined self-consistently.

In the presence of a finite magnetic field, $m_\alpha$ and $\lambda_\alpha$ ($\alpha = x,y,z$) can become finite,
leading to hybridization between the matter fermions ($\gamma_0$) and gauge fermions ($\gamma_{x,y,z}$),
which gaps out the Dirac cone for fields applied along the $\ba$ and $\bc$ directions.

\section{The four-sublattice duality transformation}\label{sec:T_4}
In the $K$-$J_3$ model, there exists a four-sublattice duality transformation that maps $(K,J_3) \rightarrow (-K,J_3)$~\cite{Rousochatzakis2024-td}.
The four sublattices are labeled by gray ($1$), red ($2$), green ($3$), and blue ($4$) in \cref{fig:T_4}.
The transformation of spin operators on each sublattice under $\scT_4$ is summarized in \cref{tab:T_4}.
\begin{table}
    \centering
    \caption{Transformation of spin operators under $\scT_4$.}
    \label{tab:T_4}
    \begin{tabular*}{1 \linewidth}{@{\extracolsep{\fill}} c c c c}
      \hline \hline
         \text{Sublattice} & $\sigma_{i}^x$ & $\sigma_{i}^y$ & $\sigma_{i}^z$  \\
      \hline
      1  & $\sigma_{1}^x$ & $\sigma_{1}^y$ & $\sigma_{1}^z$ \\
      2  & $-\sigma_{2}^x$ & $\sigma_{2}^y$ & $-\sigma_{2}^z$ \\
      3  & $-\sigma_{3}^x$ & $-\sigma_{3}^y$ & $\sigma_{3}^z$ \\
      4  & $\sigma_{4}^x$ & $-\sigma_{4}^y$ & $-\sigma_{4}^z$ \\
      \hline \hline
    \end{tabular*}
\end{table}

The existence of the $\scT_4$ duality implies that the critical values of $J_3$ for the onset of magnetic order are identical 
in the FM and AFM Kitaev models, with the corresponding ordered phases related by $\scT_4$. 
Our mean-field plus RPA calculations of the spin dynamical structure factor confirm this expectation.

\cref{fig:T_4}(a) illustrates a representative zigzag order with wave vector $\M_2$ and its $\scT_4$-transformed counterpart, the phase
with coexisting AFM and zigzag orders. 
In the latter, the spin $x$ components form a N\'eel AFM pattern, while the spin $y$ and spin $z$ components form zigzag patterns with different wave vectors.
\cref{fig:T_4}(b) shows a stripe order with wave vector $\M_2'$ and the corresponding $\scT_4$-transformed phase with coexising
FM and stripe orders.
In this case, the spin $x$ components form a uniform FM pattern, whereas the spin $y$ and spin $z$ components form stripe patterns with different wave vectors.
It should be noted that, from energetic considerations, the spin $z$ components in the zigzag (with an FM Kitaev interaction) and stripe orders
(with an AFM Kitaev interaction) shown in \cref{fig:T_4} are expected to vanish.

\section{Matrix elements in the $\scR_{J_3}$-induced Raman response}\label{sec:Raman-details}
In this section, we derive the expression for the matter-fermion matrix elements appearing in \cref{eq:C4v,eq:C2v_xx,eq:C2v_xy,eq:C2v_yx,eq:C2v_yy}.

First, we discuss those from $C^\fourv$. 
The Hamiltonian $H^a_0 + V^\fourv$ can be represented in terms of the Bogoliubov quasiparticle
$\alpha_{i}$ of $H^a_0$:
\begin{align}
    & H^a_0 + V^\fourv = \frac{1}{2} (\alpha_1^\dagger, \dots, \alpha_1, \dots) h^\fourv
    \begin{pmatrix}
        \alpha_1 \\
        \vdots \\
        \alpha_1^\dagger \\
        \vdots
    \end{pmatrix}, \\
    & h^\fourv =
    \begin{pmatrix}
        u & v^{*} \\
        v & u^{*}
    \end{pmatrix}
    \begin{pmatrix}
        \epsilon^\fourv_1 &  &  &  \\
         & \ddots & & \\
         & & -\epsilon^\fourv_1 & \\
         & & & \ddots
    \end{pmatrix}
    \begin{pmatrix}
        u^{\dagger} & v^{\dagger} \\
        v^{T} & u^{T}
    \end{pmatrix}, \\
     & (\alpha_{1,\fourv}^\dagger, \dots, \alpha_{1,\fourv}, \dots) =  (\alpha_1^\dagger, \dots, \alpha_1, \dots)
     \begin{pmatrix}
         u \\
         v
     \end{pmatrix}.
\end{align}
Because the ground state $|\Psi^a_{0,\fourv} \rangle$ of $H^a_0 + V^\fourv$ has the same fermion parity as $|\Psi^a_0 \rangle$,
$| \Psi^a_{0,\fourv} \rangle$ can be written as a BCS state of $\alpha_i$~\cite{Robledo2009-ze}:
\begin{align}
    |\Psi^a_{0,\fourv} \rangle & = \sqrt{|\det(u)|} \exp(\frac{1}{2}\sum_{i,j} F_{i,j} \alpha_i^\dagger \alpha_{j}^\dagger) | \Psi^a_0 \rangle, \\
    F & = v^* (u^*)^{-1} = -F^T.
\end{align}
In calculating the matrix element of an operator $O$ between states $|\Psi^a_0 \rangle$
and $\exp(\frac{1}{2}\sum_{i,j} F_{i,j} \alpha_i^\dagger \alpha_{j}^\dagger) |\Psi^a_0 \rangle$, it is useful to define
\begin{align}
    \langle O \rangle \equiv & \langle \Psi^a_0 | O \exp(\frac{1}{2}\sum_{i,j} F_{i,j} \alpha_i^\dagger \alpha_{j}^\dagger) | \Psi^a_0 \rangle/\mathcal{Z},
    \nonumber \\
    \mathcal{Z} = & \langle \Psi^a_0 |\exp(\frac{1}{2}\sum_{i,j} F_{i,j} \alpha_i^\dagger \alpha_{j}^\dagger) | \Psi^a_0 \rangle = 1.
\end{align}
There is then
\begin{align}
    \langle \begin{pmatrix}
        \alpha_1^\dagger \\
        \vdots \\
        \alpha_1  \\
        \vdots
    \end{pmatrix}
    ( \alpha_1^\dagger, \dots, \alpha_1, \dots ) \rangle
    = 
    \begin{pmatrix}
        0 & 0 \\
        \mathbbm{1}_N & -F
    \end{pmatrix}. \label{eq:fermion-expect}
\end{align}

Expressing $\alpha_{i,\fourv}^\dagger$ as a linear combination of $\alpha_{i}$ and $\alpha_i^\dagger$,
and then applying \cref{eq:fermion-expect}, one obtains
\begin{align}
    & \langle \Psi^a_0 | \alpha_{i,\fourv}^\dagger \alpha_{j,\fourv}^\dagger | \Psi^a_{0,\fourv} \rangle \nonumber \\
    & = \sqrt{|\det(u)|} ( u_{1,i}, \dots, u_{N,i}, v_{1,i}, \dots, v_{N,i}) \nonumber \\
    & \begin{pmatrix}
        0 & 0 \\
        \mathbbm{1}_N & -F
    \end{pmatrix}
    \begin{pmatrix}
        u_{1,j} \\
        \vdots \\
        u_{N,j} \\
        v_{1,j} \\
        \vdots \\
        v_{N,j}
    \end{pmatrix}. \label{eq:fermion-bilinear-4v}
\end{align}

Next, we discuss the matrix elements in $C^\twov$. As an example, we consider $H^a_0 + V^\twov_1$; the results for $H^a_0 + V^\twov_2$
can be obtained in the same way.
For brevity, we denote the ground state of $H^a_0 + V^\twov_1$ by $|\Psi^a_{0,\twov} \rangle$, and its Bogoliubov quasiparticle
by $\alpha_{i,\twov}$.
Because the ground state $| \Psi^a_{0,\twov} \rangle$ has opposite fermion parity relative to $| \Psi^a_0 \rangle$,
it cannot be written as a BCS state of $\alpha_i$. 
By contrast, any single-quasiparticle excited state $\alpha_{i,\twov}^\dagger | \Psi^a_{0,\twov} \rangle$
has the same parity as $|\Psi^a_0 \rangle$ and therefore can be represented as a BCS state of $\alpha_i$.
Here we take the first excited state $|\Psi^a_{1,\twov} \rangle \equiv \alpha_{1,\twov}^\dagger |\Psi^a_{0,\twov} \rangle$ as a reference
state and express it as BCS state of $\alpha_i$. Any single-quasiparticle state can then be written as
$\alpha_{l,\twov}^\dagger |\Psi^a_{0,\twov} \rangle = \alpha_{l,\twov}^\dagger \alpha_{1,\twov} | \Psi^a_{1,\twov} \rangle$ ($l=1,\dots,N$).

$H^a_0 + V^\twov$ can be written as
\begin{align}
    & H^a_0 + V^\twov = \frac{1}{2} (\alpha_1^\dagger, \dots, \alpha_1, \dots) h^\twov
    \begin{pmatrix}
        \alpha_1 \\
        \vdots \\
        \alpha_1^\dagger \\
        \vdots
    \end{pmatrix}, \\
    & h^\twov =
    \begin{pmatrix}
        \tilde{u} & \tilde{v}^{*} \\
        \tilde{v} & \tilde{u}^{*}
    \end{pmatrix}
    \begin{pmatrix}
        \epsilon^\twov_1 &  &  &  \\
         & \ddots & & \\
         & & -\epsilon^\twov_1 & \\
         & & & \ddots
    \end{pmatrix}
    \begin{pmatrix}
        \tilde{u}^{\dagger} & \tilde{v}^{\dagger} \\
        \tilde{v}^{T} & \tilde{u}^{T}
    \end{pmatrix}, \label{eq:h^2v} \\
     & (\alpha_{1,\twov}^\dagger, \dots, \alpha_{1,\twov}, \dots) =  (\alpha_1^\dagger, \dots, \alpha_1, \dots)
     \begin{pmatrix}
         \tilde{u} \\
         \tilde{v}
     \end{pmatrix}.
\end{align}

In order to represent $|\Psi^a_{1,\twov} \rangle$ as a BCS state of $\alpha_i$, one needs to construct a new BdG Hamiltonian $H'$
whose ground state is $|\Psi^a_{1,\twov} \rangle$.
This can be done by simply replacing $\epsilon^{\twov}_1 \rightarrow -\epsilon^\twov_1$
in \cref{eq:h^2v}, which gives rise to
\begin{align}
    H' = & \frac{1}{2} (\alpha_1^\dagger, \dots, \alpha_1, \dots)
    \begin{pmatrix}
        \tilde{u}' & \tilde{v}'^* \\
        \tilde{v}' & \tilde{u}'^*
    \end{pmatrix}
    \begin{pmatrix}
        \epsilon^\twov_1 & & & \\
         & \ddots & & \\
         & & -\epsilon^\twov_1 & \\
         & & & \ddots
    \end{pmatrix} \nonumber \\
    & \begin{pmatrix}
        \tilde{u}'^\dagger & \tilde{v}'^\dagger \\
        \tilde{v}'^T & \tilde{u}'^T
    \end{pmatrix}
    \begin{pmatrix}
        \alpha_1 \\
        \vdots \\
        \alpha_1^\dagger \\
        \vdots
    \end{pmatrix}.
\end{align}
The $\tilde{u}'$ and $\tilde{v}'$ are related to $\tilde{u}$ and $\tilde{v}$ through
\begin{align}
    \tilde{u}' & =
    \begin{pmatrix}
        \tilde{v}_{1,1}^* & \tilde{u}_{1,2} & \dots & \tilde{u}_{1,N} \\
        \vdots & \vdots & \ddots & \vdots \\
        \tilde{v}_{N,1}^* & \tilde{u}_{N,2} & \dots & \tilde{u}_{N,N}
    \end{pmatrix}, \\
    \tilde{v}' & =
    \begin{pmatrix}
        \tilde{u}_{1,1}^* & \tilde{v}_{1,2} & \dots & \tilde{v}_{1,N} \\
        \vdots & \vdots & \ddots & \vdots \\
        \tilde{u}_{N,1}^* & \tilde{v}_{N,2} & \dots & \tilde{v}_{N,N}
    \end{pmatrix}.
\end{align}

As the ground state of $H'$, $| \Psi^a_{1,\twov} \rangle$ can be written as
\begin{align}
    |\Psi^a_{1,\twov} \rangle & \equiv \alpha_{1,\twov}^\dagger | \Psi^a_{0,\twov} \rangle \nonumber \\
    & = \sqrt{|\det(\tilde{u}')|} \exp(\frac{1}{2}\sum_{i,j} \tilde{F}'_{i,j} \alpha_i^\dagger \alpha_j^\dagger ) |\Psi^a_0 \rangle, \\
    \tilde{F}' & = \tilde{v}'^* (\tilde{u}'^*)^{-1}.
\end{align}
Using \cref{eq:fermion-expect}, with the replacement $F \rightarrow \tilde{F}'$, one obtains
\begin{align}
    & \langle \Psi^a_0| \alpha_{l,\twov}^\dagger |\Psi^a_{0,\twov} \rangle  \nonumber \\
    & = \langle \Psi^a_0| \alpha_{l,\twov}^\dagger \alpha_{1,\twov} |\Psi^a_{1,\twov} \rangle \nonumber \\
    & = \sqrt{|\det(\tilde{u}')|}
    ( \tilde{u}_{1,l}, \dots, \tilde{u}_{N,l}, \tilde{v}_{1,l}, \dots, \tilde{v}_{N,l} ) \nonumber \\
    & \begin{pmatrix}
        0 & 0 \\
        \mathbbm{1}_N & \tilde{F}'
    \end{pmatrix}
    \begin{pmatrix}
        \tilde{v}_{1,1}^* \\
        \vdots \\
        \tilde{v}_{N,1}^* \\
        \tilde{u}_{1,1}^* \\
        \vdots \\
        \tilde{u}_{N,1}^*
    \end{pmatrix}.
\end{align}
The multi-fermion matrix elements in \cref{eq:C2v_xy,eq:C2v_yx} can be evaluated using a generalized Wick's theorem~\cite{Ring2004-jf}.
\begin{align} \label{eq:Wick-theorem}
    & \langle \Psi^a_0| \gamma_{i_1,0} \gamma_{i_2,0} \gamma_{i_3,0} \gamma_{i_4,0} \alpha_{l,\twov}^\dagger \alpha_{1,\twov} 
    \, \exp(\frac{1}{2}\sum_{i,j} \tilde{F}'_{i,j} \alpha_i^\dagger \alpha_j^\dagger ) |\Psi^a_0 \rangle \nonumber \\
    & = \mathcal{Z} \langle \gamma_{i_1,0} \gamma_{i_2,0} \gamma_{i_3,0} \gamma_{i_4,0} \alpha_{l,\twov}^\dagger \alpha_{1,\twov}  \rangle \nonumber \\
    & = \mathcal{Z} \, [  \langle \gamma_{i_1,0} \gamma_{i_2,0} \rangle \langle \gamma_{i_3,0} \gamma_{i_4,0} \rangle 
    \langle \alpha_{l,\twov}^\dagger \alpha_{1,\twov} \rangle \nonumber \\
    & + (-1) \langle \gamma_{i_1,0} \gamma_{i_2,0} \rangle \langle \gamma_{i_3,0} \alpha_{l,\twov}^\dagger \rangle 
    \langle \gamma_{i_4,0} \alpha_{1,\twov} \rangle + \dots ].
\end{align}
In total, there are 15 possible contractions in \cref{eq:Wick-theorem}. 
The expectation value of each fermion bilinear can be evaluated by first expressing the fermionic operators as
linear combinations of $\alpha_i$ and $\alpha_i^\dagger$, and then applying \cref{eq:fermion-expect}.

\bibliography{reference.bib}


\end{document}